\documentclass[11pt,reqno]{article}
\usepackage{amsmath}
\usepackage{amsfonts}
\usepackage{amssymb}
\usepackage{hyperref}
\usepackage{latexsym}
\usepackage[dvips]{graphicx}
\usepackage{epsf}

\allowdisplaybreaks \numberwithin{equation}{section}

\newcommand{\be}{\begin{equation}}
\newcommand{\ee}{\end{equation}}
\newcommand{\bi}{\begin{itemize}}
\newcommand{\ei}{\end{itemize}}
\newcommand{\bea}{\begin{eqnarray}}
\newcommand{\eea}{\end{eqnarray}}

\let\a=\alpha \let\b=\beta  \let\g=\gamma  \let\d=\delta
        \let\l=\lambda
\let\m=\mu    \let\n=\nu          \let\r=\rho
\let\s=\sigma      
    
\let\G=\Gamma \let\D=\Delta    
         
 \let\Y=\Upsilon 

\let\==\equiv

\newcommand{\tr}{{\rm tr}}
\newcommand{\Tr}{{\rm Tr}}

\newcommand{\cD}{\mathcal{D}}

\newcommand{\cL}{\mathcal{L}}
\newcommand{\cM}{\mathcal{M}}

\newcommand{\cO}{\mathcal{O}}
\newcommand{\cR}{\mathcal{R}}
\newcommand{\cS}{\mathcal{S}}
\newcommand{\cT}{\mathcal{T}}

\newcommand{\Pt}{\widetilde{\Phi}}
\newcommand{\Qt}{\widetilde{Q}}

\newcommand{\unit}{{\bf{1}}}

\newcommand{\half}{\tfrac{1}{2}}

\newcommand{\bb}{\bar{b}}

\newcommand{\gb}{\bar{g}}

\newcommand{\e}{\mathrm{e}}
\newcommand{\p}{\partial}

\newcommand{\Cb}{\bar{C}}
\newcommand{\Db}{\bar{D}}

\newcommand{\Rb}{\bar{R}}

\newcommand{\rmd}{{\rm d}}

\newcommand{\Dz}{\Delta}

\newcommand{\om}{\omega}

\newcommand{\Dbb}{\mathbb{D}}
\newcommand{\Kbb}{\mathbb{K}}
\newcommand{\Mbb}{\mathbb{M}}
\newcommand{\Pbb}{\mathbb{P}}
\newcommand{\Rbb}{\mathbb{R}}

\begin{document}
%------------------------------------------------------------------------------

\thispagestyle{empty}
\begin{flushright} \small
MZ-TH/10-42\\
AEI-2010-165
\end{flushright}
\bigskip

\begin{center}
 {\LARGE\bfseries   The Universal RG Machine 
}
\\[10mm]
Dario Benedetti$^1$, Kai Groh$^{3}$, Pedro F. Machado$^2$ and Frank Saueressig$^{3}$ \\[3mm]
$^1${\small\slshape
Max Planck Institute for Gravitational Physics, Albert Einstein Institute,\\ Am M\"{u}hlenberg 1, D-14476 Potsdam, Germany \\
{\upshape\ttfamily dario.benedetti@aei.mpg.de} } \\[3mm]
$^2${\small\slshape
Department of Genetics, University of Cambridge \\
Downing Street, CB2 3EH, Cambridge, UK\\
{\upshape\ttfamily p.farias-machado@gen.cam.ac.uk} }
\\[3mm]
$^3${\small\slshape
Institute of Physics, University of Mainz\\
Staudingerweg 7, D-55099 Mainz, Germany \\
{\upshape\ttfamily kgroh@thep.physik.uni-mainz.de} \\
{\upshape\ttfamily saueressig@thep.physik.uni-mainz.de} }\\

\end{center}
\vspace{5mm}

\hrule\bigskip

\centerline{\bfseries Abstract} \medskip
\noindent
Functional Renormalization Group Equations constitute a powerful tool to encode the perturbative and non-perturbative properties of a physical system.
We present an algorithm to systematically compute the expansion of such flow equations in a given background quantity specified by the approximation scheme. The method is based on off-diagonal heat-kernel techniques and can be implemented on a computer algebra system, opening access to complex computations in, e.g., Gravity or Yang-Mills theory. In a first illustrative example, we re-derive the gravitational $\beta$-functions of the Einstein-Hilbert truncation, demonstrating their background-independence. As an additional result, the heat-kernel coefficients for transverse vectors and transverse-traceless symmetric matrices are computed to second order in the curvature.
\bigskip
\hrule\bigskip
\newpage
\tableofcontents
\newpage
%------------------------------------------------------------------------------
%-------------------------------------------------------------
\section{Introduction}
\label{sect:intro}
%-------------------------------------------------------------

%-------------------------------------------------------------
\subsection{The Functional Renormalization Group}
\label{sect:intro1}
%-------------------------------------------------------------
Functional Renormalization Group Equations (FRGEs) constitute a universal and highly flexible tool for unlocking non-perturbative information in a plethora of settings ranging from condensed matter physics, to statistical and elementary particle physics, to quantum gravity \cite{Berges:2000ew,Wetterich:2001kra,Pawlowski:2005xe,Gies:2006wv,Rosten:2010vm}. They provide an explicit realization of Wilson's ideas of renormalization \cite{Wilson:1973jj}, allowing us to integrate out quantum fluctuations in a system ``shell-by-shell" in a much more manageable way than, e.g., directly manipulating the path-integral.

One of the main practical virtues of the FRGEs is that general non-perturbative properties of a theory, like phase transitions or fixed points of the renormalization group (RG) flow, can be found via relatively simple approximations. This feature makes FRGEs a primary tool in the search for non-Gaussian Fixed Points (NGFPs) of the RG flow in Quantum Field Theories, which provide a non-trivial and predictive generalization of the notion of perturbative renormalizability within the framework of Asymptotic Safety \cite{Weinberg:1976xy,ZinnJustin:2002ru}. The existence of such NGFPs has been well-established in the context of, e.g., $O(N)$-sigma-models \cite{Brezin:1975sq} and has recently also been discussed in the context of standard model physics related to Yukawa-systems \cite{Gies:2009hq,Gies:2009sv,Scherer:2009wu} and the Higgs sector \cite{Fabbrichesi:2010xy}, where it has lead to new physics scenarios beyond the ones allowed within perturbative renormalization. A NGFP has also been conjectured to control the UV behavior of four-dimensional gravity \cite{Weinberg:1980gg} (see also \cite{Weinberg:2009ca,Weinberg:2009bg}), rendering the theory non-perturbatively renormalizable or, equivalently, asymptotically safe. This conjecture is motivated by the fixed-point structure of gravity in $d=2+\epsilon$ dimensions \cite{Christensen:1978sc,Gastmans:1977ad} and evidence in its favor has been established in a series of works (see \cite{Niedermaier:2006wt,Percacci:2007sz,Reuter:2007rv,Litim:2008tt} for reviews and detailed references).

While simple approximations of the exact RG flow, usually consisting of truncating the flow to a handful of running couplings, can already yield qualitative information on the fixed-point structure of a given theory, establishing the consistency of the approximation and answering questions concerning, e.g., critical exponents, universality classes and the number of relevant couplings appearing in the theory requires more sophistication. For the case of gravity, in particular, the best truncations to date \cite{Codello:2007bd, Machado:2007ea,Narain:2009gb,Benedetti:2009rx,Benedetti:2009gn,Benedetti:2009iq} have mostly been restricted to fluctuations around a spherical background, and are insufficient to properly answer these questions. In this light, it is highly desirable to have a computational tool which can track the RG flow on a sufficiently large subspace of all running coupling constants.

The main purpose of this paper is to present such an algorithm, which we call the ``universal RG machine'', that allows us to compute approximations of the exact FRGE in a flexible and systematic way. Its key element is the algorithmic reduction of the (usually complicated) operator traces appearing in the flow equation to the off-diagonal heat-kernel expansion of a standard Laplace-operator on a curved manifold \cite{DeWitt:1975ys,Christensen:1978yd} (see also \cite{Decanini:2005eg,Anselmi:2007eq} for current results and further references). This procedure is general and does not rely on the choice of a particular background, which is usually otherwise evoked to simplify computations. Furthermore, the bookkeeping required in a practical computation can be easily handled by a computer algebra program, opening the door to computations which  until now have been out of reach due to technical limitations. The algorithm will be presented here having gravity as the underlying theory, but we stress that the construction is readily adapted to other gauge-theories.

The rest of the paper is organized as follows. In the next subsection, we give a short introduction to the Wetterich equation \cite{Wetterich:1992yh}, which provides the foundation of our construction. Sect.\ \ref{sect:algo} contains the blueprint of the universal RG machine and constitutes the main result of our work. As an illustration, we apply our construction to the  Einstein-Hilbert truncation \cite{Reuter:1996cp,Reuter:2001ag,Lauscher:2001ya,Litim:2003vp,Codello:2008vh} in Sect.\ \ref{sect:example}, before closing with an outlook on future research perspectives in Sect.\ \ref{sect:disc}. Proofs of the more technical details entering the construction have been relegated to five appendices, which cover the commutator relations, transverse-traceless decomposition of the fluctuation fields, the derivation of the heat-kernel coefficients for transverse vectors and transverse-traceless matrices, and general considerations on the structure of the flow equation regarding gauge-fixing and truncation independent contributions.

%-------------------------------------------------------------
\subsection{The Wetterich Equation}
\label{sect:intro2}
%-------------------------------------------------------------
The starting point of the universal RG machine is the FRGE for the effective average action $\Gamma_k$ \cite{Wetterich:1992yh}, which schematically takes the form
\be\label{FRGE}
\p_t \Gamma_k[\phi, \bar{\phi}] = \half {\rm STr} \left[ \left( \Gamma_k^{(2)}  + \cR_k \right)^{-1} \, \p_t \cR_k  \right]\, , \quad \p_t = k \frac{\rmd}{\rmd k} \, .
\ee
Its construction for gauge-theories \cite{Reuter:1993kw} and gravity \cite{Reuter:1996cp} employs the background field formalism, which ensures the background covariance of the flow equation and implies that $\Gamma_k[\phi, \bar \phi]$ depends on two arguments, the background fields $\bar \phi$ and the fluctuations around this background, collectively denoted by $\phi$. The Hessian $\Gamma_k^{(2)}$ is the second variation of $\Gamma_k$ with respect to the fluctuation fields $\phi_i$ at fixed background $\bar \phi$ and is, in general, matrix-valued in field space,
\be\label{Gamma2}
\left[ \Gamma^{(2)}_k\right]^{ij} = (-1)^{[j]} \, \frac{1}{\sqrt{\gb}} \, \frac{1}{\sqrt{\gb}} \,  \frac{\delta^2 \Gamma_k}{\delta \phi_i \delta \phi_j} \, , 
\ee
where $\gb$ is the determinant of the background space-time metric (which equals unity for flat space) and $[j]$ takes the values 0,1 for bosonic and fermionic fields respectively. In the FRGE, STr is a generalized functional trace which includes a minus sign for ghosts and fermions, and a factor of two for complex fields, while $\cR_k(p^2)$ is a matrix valued IR cutoff which provides a $k$-dependent mass-term for fluctuations with momenta $p^2 < k^2$.
The cutoff satisfies $\cR_k \propto k^2$ for $p^2 \ll k^2$ and vanishes for high-momentum modes, $\cR_k \rightarrow 0$ as $p^2 \gg k^2$.
The factor $\p_t \cR_k$ ensures that the r.h.s.\ of the flow equation is finite and peaked at momenta $p^2 \approx k^2$, so that an additional UV regulator becomes dispensable.
Eq.\ \eqref{FRGE} describes the dependence of $\G_k[\phi, \bar{\phi}]$ on a renormalization group scale $k$, realizing the Wilsonian picture of renormalization by integrating out quantum fluctuations with momenta $k^2 \le p^2 \le \Lambda$. Its solutions interpolate between the ordinary effective action $\Gamma[\bar{\phi}] = \Gamma_{k = 0}[\phi = 0, \bar{\phi}]$ and an initial action $\Gamma_\Lambda$ at the UV scale, which in the limit $\Lambda \rightarrow \infty$ essentially reduces to the bare action (see \cite{Manrique:2009uh} for more details).

The main shortcoming of the FRGE is that it cannot be solved exactly. In order to extract physics from it, one thus has to resort to approximations. In the one-loop approximation, where $\Gamma_k$ on the r.h.s. of \eqref{FRGE} is replaced by the $k$-independent bare action, perturbative results are reproduced. Going beyond perturbation theory, a common strategy relies on expanding $\Gamma_k[\phi, \bar{\phi}]$ on a basis of local invariants and truncating the series, thereby limiting the form of the effective average action to a finite number of invariants $I_n[\phi, \bar{\phi}]$,
\be\label{trunc}
\Gamma_k[\phi, \bar{\phi}] \approx \sum_{n=1}^N \, u_n(k) \, I_n[\phi, \bar{\phi}] \, .
\ee
Substituting such ansatz into \eqref{FRGE} and projecting the exact RG flow onto the subspace spanned by the $I_n[\phi, \bar{\phi}]$ then provides the $\beta$-functions for the $k$-dependent couplings $u_n(k)$. As this approximation scheme does not rely on an expansion in a small parameter, the resulting $\beta$-functions typically capture contributions from all loop-orders and constitute non-perturbative results. Possible ordering principles for organizing the expansion \eqref{trunc} include, e.g., the derivative expansion, whereby interactions with increasing number of derivatives are systematically included in the ansatz, or a vertex expansion, in which \eqref{trunc} is truncated at a certain order of the fluctuation field $\phi$.

Depending on the complexity of the ansatz \eqref{trunc}, projecting the right-hand side of the flow equation onto the $I_n[\phi, \bar{\phi}]$ can become quite involved, since the matrix-valued operator $( \Gamma_k^{(2)}  + \cR_k )$ is usually of non-minimal form, which poses a severe complication for the evaluation of the functional trace in \eqref{FRGE}. This obstacle is frequently circumvented by working with a particular background configuration, with the drawback that some $I_n[\phi, \bar{\phi}]$ become indistinguishable, thus potentially hiding important features of the theory under investigation. In the gravitational context, for example, it is common to restrict the background metric to be maximally symmetric, i.e., a sphere. In such a case the only terms that can be recognized are integral powers of the Ricci scalar, and it is not possible to distinguish other important invariants, such as  the terms $I = \int d^dx \sqrt{g} R^{\m\n}R_{\m\n}$ or $I = \int d^dx \sqrt{g} R D^2 R$, which can affect the counting of the degrees of freedom of the theory.
In order to distinguish such invariants, we need a generic background, but this leads to serious technical complications due to the operator structure appearing in the second variation of the action.
It is thus clear that a new computational scheme, enabling us to deal with the general operator structures appearing in the FRGE without resorting to a simplifying choice of background, is highly desirable. In the next section we will propose such an algorithm in form of the  universal RG machine.

%-------------------------------------------------------------
\section{The universal renormalization group machine}
\label{sect:algo}
%-------------------------------------------------------------
We will start by discussing the universal RG machine from a general perspective, before giving an explicit example of its implementation in the case of gravity in Sect.\ \ref{sect:example}. The algorithm that we propose consists of the following steps:
\begin{center}
\begin{tabular}{l p{12cm}}
1. & expanding $\G_k$ to quadratic order in the fluctuation fields $\phi_i$; \\
2. & simplifying the operator structure of $\Gamma_k^{(2)}$ by performing the transversal decompositions \eqref{vec:dec} and \eqref{mTTdec} of the vector and tensor fluctuations, respectively; \\
3. & decomposing $\Gamma_k^{(2)}$ into its kinetic part $\Kbb$ and interactions $\Mbb$ according to \eqref{op:class}; \\
5. & inverting $[\G_k^{(2)}+\cR_k]$ via the perturbative expansion in the $\Mbb$'s, eq.\ \eqref{exp1}; \\
6. & inserting projection operators in the traces; \\
7. & bringing the trace arguments into the canonical form \eqref{cantrace} by applying commutator relations for covariant derivatives; \\
8. & evaluating traces without interactions via standard heat-kernel expansions for transverse fields; \\
9. & replacing derivatives appearing in $\Mbb$ by the $H$-tensors \eqref{anselmitensor} constructed from the off-diagonal heat-kernel; \\
10. & reading off $\beta$-functions by matching background structures on both sides of \eqref{FRGE} ; 
\end{tabular}
\end{center}
Notably, each of these steps can be handled by computer algebra software. This is of crucial advantage when dealing with more complex expansions (truncations),
since these require an increasing amount of bookkeeping when collecting the various contributions to the RG flow. We now proceed by discussing the steps indicated above in more detail.
%
%-------------------------------------------------------------
\subsection{Simplifying the operator structure of $\Gamma_k^{(2)}$}
%-------------------------------------------------------------
Schematically, $\Gamma_k^{(2)}$ can be written as
\be\label{op:class}
\left[ \Gamma^{(2)}_k \right]^{ij} = \qquad \underbrace{\Kbb_i(\Delta) \, \delta^{ij} \, \unit_i}_{\rm kinetic\ terms} \qquad + \underbrace{\mathbb{D}^{ij}(D_\m)}_{\rm uncontracted\ derivatives} + \quad \underbrace{\mathbb{M}^{ij}(R, D_\m)}_{\rm background\ curvature} \, ,
\ee
where $i,j$ denote the indices in field space, $\unit_i$ denotes the unit of the corresponding field space, and $\Delta = - g^{\m\n} D_\m D_\n$ is the Laplace-operator. The first piece, $\Kbb_i(\Delta)$, is the diagonal part of the inverse propagator of the $i$-th field, containing only Laplacians and eventually cosmological and mass terms.
The last two pieces group the off-diagonal operators, with $\mathbb{M}^{ij}$ including all interactions that involve at least one power of the object controlling the expansion of the flow equation \eqref{trunc}, and with $\mathbb{D}^{ij}$ containing all the remaining terms with derivatives but no such expansion object. For a single metric truncation of the flow equation, the expansion field is naturally given by the curvature tensor of the background metric. The formalism, however, allows for an expansion in any background structure like the background ghost fields $c_\n$ or, in the case of a vertex expansion, the fluctuation fields $\phi$. 
Explicit examples of typical off-diagonal matrix entries are
\be\label{D-example}
\mathbb{D} = (1-\alpha) D_\m D^\n \, , \qquad \mathbb{D} = D^\m D^\n D_\alpha D_\beta \, ,
\ee
and
\be
\mathbb{M} = R^{\m\n} D_\m D_\n \, , \qquad \mathbb{M} =  D^\m c_\n \, .
\ee
Our goal is to evaluate the functional trace on the r.h.s. of \eqref{FRGE} by means of off-diagonal heat-kernel techniques. The first step to this aim is to simplify the operator structure of \eqref{op:class} so that terms of the $\Dbb$-type do not occur. The reason for this is that operators of this kind would contribute with arbitrary powers, spoiling the perturbative inversion of the modified propagator $\left[ \Gamma_k^{(2)} + \cR_k \right]$ which we will subsequently employ.

One way to remove these operators, often used in connection to ordinary perturbation theory, is to choose a convenient gauge-fixing condition. Considering the first example in \eqref{D-example}, for instance, the corresponding $\Dbb$-term could be removed by choosing the value $\alpha = 1$ for the gauge parameter. Following this route is problematic for two reasons, however. Firstly, it restricts the gauge-sector to a very particular choice, which has to be carefully adapted to the truncation to be studied. Secondly, in most cases there will not be enough freedom in the gauge-fixing function to remove all $\Dbb$-type terms. The latter is the case for, e.g., higher-derivative gravity, where conveniently-chosen gauges can remove all four-derivative $\Dbb$-terms, but lower-derivative terms still remain \cite{Barth:1983hb}.

A systematic way to remove these terms is the transverse-traceless- (TT) (or York-) decomposition of the fluctuation fields on a generic gravitational background. For vector fluctuations this decomposition is given by
\be\label{vec:dec}
\phi_\m = \phi_\m^{\rm T} + D_\m \eta\,,
\ee
subject to the differential constraint
\be\label{vec:dconst}
D^\m \phi_\m^{\rm T} = 0 \, .
\ee
Analogously, metric fluctuations are split according to the ``minimal'' TT-decomposition\footnote{The ``standard'' TT-decomposition further decomposes the vector $\xi_\m$ according to \eqref{vec:dec}, as is explicitly done in App.~\ref{App:E}. %
} 
\be\label{mTTdec}
h_{\m\n} = h_{\m\n}^{\rm T} + D_\m \xi_\n + D_\n \xi_\m - \tfrac{2}{d} g_{\m\n} D^\alpha \xi_\alpha + \tfrac{1}{d} g_{\mu\nu} h \, ,
\ee
with the component fields satisfying 
\be\label{TTconst}
D^\m h_{\m\n}^{\rm T} = 0 \,  ,\qquad g^{\m\n} h^{\rm T}_{\m\n} = 0 \, , \qquad g^{\m\n} h_{\m\n} = h \, .
\ee
 At the price of introducing additional fields, the transverse constraints on the fluctuation fields guarantee that any $\Dbb$-type term can be converted into $\Mbb$-contributions. This follows from the fact that at least one covariant derivative contained in $\Dbb$ has to be contracted with a transverse fluctuation field. Using standard commutation relations, this derivative can then be moved to the very left or very right of $\Dbb$ so that it acts on the fluctuation in such a way that the resulting term vanishes. In the process, one picks up additional terms which contain background curvatures and are therefore of $\Mbb$-type by definition. As a result, after performing the TT-decomposition of the metric fluctuations and rearranging the resulting operators, we end up with no $\Dbb$-terms in $\Gamma_k^{(2)}$.

%-------------------------------------------------------------
\subsection{Constructing the inverse of $\left[ \Gamma_k^{(2)} + \cR_k \right]$}
%-------------------------------------------------------------
Having obtained the inverse propagator $\Gamma_k^{(2)}$ in simplified form, the universal RG machine proceeds with the construction of the regulated propagator $\left[ \Gamma_k^{(2)} + \cR_k \right]^{-1}$ by expanding the latter in the background structure controlling the truncation, as, e.g. the background curvature. The central idea here is to treat the off-diagonal, $\Mbb$-type operators appearing in the inverse regulated propagator as perturbations on the diagonal, Laplace-type operators in $\Kbb$. The presence of the expansion field in the $\Mbb$-terms  allows us to perturbatively invert  $\left[ \Gamma_k^{(2)} + \cR_k \right]$ in $\Mbb$ along the same expansion scheme as that of the effective average action, thereby accounting for all contributions to any given truncation.

%-------------------------------------------------------------
\subsubsection{Adapting the infrared cutoff}
%-------------------------------------------------------------
Following standard FRGE practice, we first modify the inverse propagator $\Gamma_k^{(2)}$ by a cutoff $\cR_k$. With the structure of the inverse propagator split into $\Kbb$ and $\mathbb{M}$, it is natural to tailor the cutoff to the $\Kbb$-term alone. Following this scheme, the Laplacians appearing in $\Kbb$ are dressed up by the momentum-dependent IR regulator, according to the rule
\be\label{typeIcutoff}
\Delta \mapsto P_k = \Delta + R_k(\Delta) \, ,
\ee
where $R_k(\Delta)$ denotes the basic (scalar) cutoff function, which interpolates monotonously between $R_k(z) \propto k^2, z \ll k^2$ and $R_k(z) = 0, z \gg k^2$.
As a consequence, the matrix-valued kinetic terms in the modified inverse propagator take the form
\be\label{TypeIb}
\Kbb(\Delta) \mapsto \Pbb(\Delta) = \Kbb(\Delta) + \Rbb_k(\Delta) = \Kbb( P_k )\,,
\ee
where $\Pbb$ is obtained from $\Kbb$ by applying the replacement rule \eqref{typeIcutoff}. This construction guarantees that the IR-regulator $\cR_k$ is diagonal in field space. Note that $\Kbb$, and therefore also $\Pbb$ and $\Rbb_k$, will in general contain $k$-dependent coupling constants such as wave-function renormalization factors.

It is useful to compare our construction with cutoff scheme classification of \cite{Codello:2008vh}. Since \eqref{typeIcutoff} regulates the Laplacian only, this scheme falls into the class of Type I cutoffs. However, since only Laplace-operators appearing in $\Kbb$ are regulated (and not those that could occur in $\Mbb$) this is not the standard Type I cutoff, but a variation of it which we will call ``Type Ic" scheme in the following. We note that the Type I and Ic cutoffs agree for truncations containing only two space-time derivatives, while they will generically differ for higher-derivative truncations.

%-------------------------------------------------------------
\subsubsection{Inverting the regulated inverse propagator}
%-------------------------------------------------------------
Having completed the cutoff implementation, the inverse regulated propagator becomes a block-matrix in field space
\be\label{G2structure}
\left[ \Gamma^{(2)} + \cR_k \right]^{ij} = \left[
\begin{array}{cc}
\Pbb_A(\Delta) \, \unit_A + \Mbb_A & \Mbb_\times \\
\Mbb_\times^\dagger & \Pbb_B(\Delta) \, \unit_B + \Mbb_B
\end{array}
\right]\,.
\ee
While we restricted ourselves to the case of two fields here, as this is what is typically encountered in practical computations, the generalization to the case of more than two interacting fields is straightforward.

In order to obtain the modified propagator from \eqref{G2structure}, we first formally apply the exact inversion formula for block-matrices,
\be\label{Ginv}
\left[\begin{array}{cc} A & B \\ C & D \end{array}\right]^{-1} = \left[\begin{array}{cc} \left(A - B D^{-1} C\right)^{-1} & - A^{-1}B\left(D - C A^{-1} B \right)^{-1} \\ -D^{-1} C \left(A - BD^{-1} C\right)^{-1} & \left(D - C A^{-1} B\right)^{-1} \end{array}\right]\,.
\ee
Building on the split between the kinetic terms $\Pbb$ and the interaction terms $\Mbb$, each block-matrix entry above can be constructed {\it perturbatively} in $\Mbb$.\footnote{We stress that this perturbative inversion should not be confused with standard perturbation theory, where the power series is controlled by a small coupling constant. Here, the perturbative series is controlled by a background quantity, the expansion field, which organizes the truncation of the effective action. Consequently, this approach still allows us to work ``non-perturbatively'' in the coupling constants.}
This computation may be simplified further by observing that the cutoff operator $\cR_k$ and thus $\p_t \cR_k$ are diagonal in field space. Thus, it suffices to consider the diagonal elements in \eqref{Ginv}. Substituting for \eqref{G2structure}, the perturbative inversions in $\Mbb$ for the diagonal elements read
\be\label{exp1}
\begin{split}
\left[ \Gamma^{(2)} + \cR_k \right]_{AA}^{-1} = & \, \frac{1}{\Pbb_A} - \frac{1}{\Pbb_A} \, \Mbb_A  \, \frac{1}{\Pbb_A} + \frac{1}{\Pbb_A} \, \Mbb_A \, \frac{1}{\Pbb_A} \, \Mbb_A \, \frac{1}{\Pbb_A}
+ \frac{1}{\Pbb_A} \, \Mbb_\times \, \frac{1}{\Pbb_B} \, \Mbb_\times^\dagger \, \frac{1}{\Pbb_A}
+\cO(\Mbb^3)\, , \\
\left[ \Gamma^{(2)} + \cR_k \right]_{BB}^{-1} = & \, \frac{1}{\Pbb_B} - \frac{1}{\Pbb_B} \, \Mbb_B  \, \frac{1}{\Pbb_B} + \frac{1}{\Pbb_B}  \, \Mbb_B  \, \frac{1}{\Pbb_B} \, \Mbb_B \, \frac{1}{\Pbb_B}
+ \frac{1}{\Pbb_B}  \, \Mbb_\times^\dagger \, \frac{1}{\Pbb_A} \, \Mbb_\times \, \frac{1}{\Pbb_B}
+\cO(\Mbb^3)\, ,
\end{split}
\ee
where the $\Mbb_i$ are understood as suitably projected onto the subspace spanned by the fluctuation fields. As $\Mbb$ contains the background quantity organizing the expansion scheme of the flow equation, we are ensured that only a finite number of terms appearing in these expansions contribute to a given truncation.

%-------------------------------------------------------------
\subsection{Evaluating the traces via off-diagonal heat-kernel expansion}
\label{sect:2.3}
%-------------------------------------------------------------
Substituting the perturbative expansion of $\left[ \Gamma_k^{(2)} + \cR_k \right]^{-1}$ into the r.h.s.\ of the flow equation results in a series of individual operator traces on a space of fields satisfying the differential constraints imposed by the TT-decomposition. We will now discuss the evaluation of these traces, distinguishing them into two different classes according to whether or not they contain background vertices $\Mbb$.

%-------------------------------------------------------------
\subsubsection{Traces without non-minimal operator insertions}
%-------------------------------------------------------------
The traces without background vertices typically assume the simple form
\be\label{trace:kin}
\cS^{\rm kin} = \Tr_{s} \left[ \frac{1}{\Pbb} \p_t \Rbb_k \right]\,,
\ee
where $s$ indicates the ``spin'' of the fluctuation field and $s = 0, 1, 2, {\rm 1T}, {\rm 2T}$ denotes the trace with respect to a scalar, vector, symmetric tensor, or a vector and symmetric tensor satisfying transverse-traceless conditions, respectively. As this class of traces contains only Laplace operators in their arguments, they can be evaluated via standard heat-kernel methods by expressing the operator occurrence as a Laplace-transform and then evaluating a standard heat trace of the form
\be\label{hexp}
\Tr_{s} \left[ \e^{-s \Delta} \right] = \frac{1}{(4 \pi s)^{d/2}} \int d^dx \sqrt{g} \left[ c_1 + c_2 R + c_3 R^3 + c_4 R_{\m\n} R^{\m\n} + c_5 R_{\m\n\a\b} R^{\m\n\a\b} \right] .
\ee

For fields without differential constraints, these coefficients are well-known and can be found in, e.g., \cite{Vassilevich:2003xt,Avramidi:2000bm}. For fields satisfying differential constraints, ${\rm 1T}$ and ${\rm 2T}$ coefficients are known for the special classes of maximally symmetric backgrounds \cite{Lauscher:2001ya} or Lichnerowicz-Laplacians on Einstein-spaces \cite{Benedetti:2009gn} only. In Appendix \ref{sect:heat}, we have computed these heat kernel coefficients for the case of a general, purely gravitational, background. The result of this computation is summarized in Table \ref{t.heatcoeff} for $d=4$.

\begin{table}[t!]
\begin{center}
\begin{tabular}{|c|c|c|c|c|c|}\hline
 & $\qquad \quad  0 \qquad \quad$ & $\qquad \quad 1 \qquad \quad$ & $\qquad \quad 2 \qquad \quad$ & $\qquad \quad {\rm 1T} \qquad \quad $ & $\qquad \quad {\rm 2T} \qquad \quad$  \\ \hline
 $c_1$ & $1$ & $4$ & $10$ & $3$ & $5$ \\ \hline
 $c_2$ & $\frac{1}{6}$ & $\frac{2}{3}$ &  $\frac{5}{3}$ & $\frac{1}{4}$ & $-\frac{5}{6}$   \\ \hline
 $c_3$ & $\frac{1}{72}$ & $\frac{1}{18}$ &  $\frac{5}{36}$ & $-\frac{1}{24} +\frac{1}{2\chi_{\rm E}}$ & $-\frac{137}{216} +\frac{N}{2\chi_{\rm E}}$   \\ \hline
 $c_4$ & $- \frac{1}{180}$ &  $- \frac{1}{45}$ &  $-\frac{1}{18}$ & $\frac{1}{40} -\frac{2}{\chi_{\rm E}}$ & $-\frac{17}{108} -\frac{2N}{\chi_{\rm E}}$  \\ \hline
 $c_5$ & $\frac{1}{180}$ &  $- \frac{11}{180}$ &  $- \frac{4}{9}$ &  $-\frac{1}{15} +\frac{1}{2\chi_{\rm E}}$ & $\frac{5}{18} +\frac{N}{2\chi_{\rm E}}$ \\ \hline
\end{tabular}
\caption{\label{t.heatcoeff} Coefficients appearing in the heat-kernel expansion \eqref{hexp} for $d=4$. The coefficients $c_i$ for the ${\rm 1T}$ and ${\rm 2T}$-fields are novel and computed in Appendix \ref{sect:heat}. Here, the terms including the Euler characteristic $\chi_{\rm E}$ are due to zero modes of the decompositions, and $N$ is the sum of conformal and non-conformal Killing-vectors of the background manifold.}
\end{center}
\end{table}

%-------------------------------------------------------------
\subsubsection{Traces containing non-minimal operator insertions}
\label{sect:nonmin}
%-------------------------------------------------------------
The second kind of contributions to the flow equation arises from traces that contain background vertices $\Mbb$ with derivative operators of non-Laplacian form. In order to evaluate these, we first employ commutation relations for these covariant derivatives to arrange the trace arguments into ``standard form'', so that all Laplacians are collected in a single function and the product of the background vertices forms a single operator insertion $\cO$. The trace can then be written as
\be\label{cantrace}
\cS^{\rm vertex} = \Tr_{s} \left[ W(\Delta) \, \cO \right] \, .
\ee
Denoting the Laplace transform of $W(x)$ by $\tilde{W}(s)$, this trace can be re-expressed as
\be\label{tr:canform}
\Tr_i \left[ \,  W(\Delta) \,  \cO \, \right] = \int_0^\infty ds \, \tilde{W}(s) \,  \langle x_i | \,  \e^{-s \Delta} \,  \cO \,| x_i \rangle \, ,
\ee
where $|x_i\rangle$ form a basis of eigenfunctions of the Laplace-operator in position space, carrying appropriate Lorentz indices.
Inserting a complete set of states, \eqref{tr:canform} can be evaluated using off-diagonal heat-kernel methods \cite{Anselmi:2007eq}. Explicitly,
\be\label{off-diag.trace}
\begin{split}
\langle x_i | \, \cO \, \e^{-s \Delta}  | x_i \rangle = & \, \langle x_i | \, \cO \, |x^\prime_i \rangle \langle x^\prime_i| \e^{-s \Delta}  | x_i \rangle
= \int d^4x \sqrt{g} \, {\rm tr}_i \left[ \cO H(s, x, x^\prime) \right]_{x = x^\prime} \, ,
\end{split}
\ee
where $H(x, x^\prime; s)$ is the heat-kernel at non-coincident points
\be
H(x, x^\prime; s) := \langle x^\prime | \e^{-s \Dz} | x \rangle = (4 \pi s)^{-d/2} \, \e^{- \frac{\sigma(x,x^\prime)}{2 s}} \, \sum_{n=0}^\infty s^n A_{2n}(x, x^\prime) \, .
\ee
Here, $\sigma(x,x^\prime)$ denotes half the squared geodesic distance between $x$ and $x^\prime$ and $A_{2n}(x, x^\prime)$, $n \in \mathbb{N}$, are the off-diagonal heat-kernel coefficients. The latter are  purely geometrical two-point objects, formally independent of the space-time dimension $d$, \footnote{For some early work on the spin-dependence of the $A_n(x, x^\prime)$ see also  \cite{Christensen:1978yd}.} and explicit values for them can be found in, e.g., \cite{Decanini:2005gt} and references therein.

In practice, the explicit evaluation of the trace \eqref{off-diag.trace} requires that the covariant derivatives contained in $\cO$ and acting on $H(x, x^\prime; s)$ be totally symmetrized in their indices. This can be achieved by successively writing non-symmetric combinations of the derivatives as a sum of symmetric and antisymmetric pieces and expressing the latter in terms of curvatures using the commutation rule \eqref{basiscom}. The coincidence limit of the tensors arising after acting with the symmetrized derivatives on $H(x, x^\prime; s)$ can be computed making use of the following properties of $\sigma(x, x^\prime)$ and $A_{2n}(x, x^\prime)$. Denoting the coincidence limit $x \rightarrow x^\prime$ by an overline and using round brackets for symmetrization of indices with unit strength, $(\a\b) = \half(\a\b +\b\a)$, etc, we have
\be
\half \sigma^{;\mu} \sigma_{;\mu} = \sigma \, , \qquad \overline{\sigma(x,x^\prime)} = \overline{\sigma(x,x^\prime)_{;\mu}} = 0 \, , \qquad  \overline{\sigma(x, x^\prime)_{;\mu(\alpha_1 \ldots \alpha_n)}} = 0 \, , \; n \ge 2 \, .
\ee
The only non-trivial contribution from $\sigma(x, x^\prime)$ arises when exactly two covariant derivatives act symmetrically on it,
\be
\overline{\sigma(x,x^\prime)_{;(\mu\nu)}} = g_{\m\n}(x) \, .
\ee
Secondly, up to terms of $\cO(R^2)$, the contributions from derivatives acting on the off-diagonal heat-kernel coefficients $A_{2n}(x, x^\prime)$ become
\be
\overline{A_0(x, x^\prime)} = 1 \, , \quad \overline{A_{0}(x, x^\prime)_{;\mu}} = 0 \, , \quad \overline{A_{0}(x, x^\prime)_{;(\mu\nu)}} = \frac{1}{6} R_{\mu\nu} \, , \quad \overline{A_2(x, x^\prime)} = \frac{1}{6} R \, .
\ee
Using these relations and defining
\be
H_{\a_1 \ldots \a_{2n}} := \overline{D_{(\a_1} \cdots D_{\a_{2n})} \, H(x, x^\prime; s) } \, ,
\ee
we arrive at the following expansions, valid up to terms of $\cO(R^2)$ and given here for later reference,
\be\label{anselmitensor}
\begin{split}
H_{\a_1 \ldots \a_{2n}}  = &\, \frac{1}{(4\pi s)^{d/2}} \bigg\{  \left( - 2 s \right)^{-n} \Big( g_{\a_1 \a_2} \cdots g_{\a_{(2n-1)} \a_{2n}} + (\tfrac{(2n)!}{2^n n!} - 1) \; \mbox{perm.} \Big) \Big( 1 + \tfrac{1}{6} s R \Big) \\
& \, + \tfrac{1}{6} \left( - 2 s \right)^{-(n-1)} \Big( g_{\a_1 \a_2} \cdots g_{\a_{2n-3} \a_{2n-2}} R_{\alpha_{2n-1} \alpha_{2n}} + (\tfrac{(2n)!}{2^n (n-1)!} - 1) \; \mbox{perm.} \Big)
\bigg\} \, .
\end{split}
\ee
Explicitly,
\be\label{atexplicit}
\begin{split}
H_{\a\b} = & \, \frac{1}{(4\pi s)^{d/2}} \Big\{ -  \frac{1}{2s} \, g_{\a\b} \,  \left(1 + \tfrac{1}{6} s R \right) + \frac{1}{6} R_{\a\b} \Big\} \, , \\
H_{\a\b\m\n} = & \, \frac{1}{(4\pi s)^{d/2}} \Big\{ \frac{1}{4s^2} \left( g_{\a\m} g_{\b\n} + g_{\a\n} g_{\b\m} + g_{\a\b} g_{\m\n} \right) \left(1 + \tfrac{1}{6} s R \right) \\
& \qquad - \frac{1}{12s} \left( g_{\a\m} R_{\b\n} + g_{\a\n} R_{\b\m} + g_{\b\m} R_{\a\n} + g_{\b\n} R_{\a\m} + g_{\a\b} R_{\m\n} + g_{\m\n} R_{\a\b} \right)
\Big\} \, .
\end{split}
\ee
For a given truncation, the types of $H$-tensors required can be determined by observing that each power of the curvature contained in $\Mbb$ can be contracted with at most two covariant derivatives. To a given order $n$ in the curvature, their total occurring number is therefore given by $2n$ for a scalar trace, whereas for traces over (unconstrained) vector fields and symmetric tensors, the number of required derivatives increases by 2 and 4 respectively, since they carry open indices. This number is, of course, modified if the background structure used in the expansion carries a different set of indices.

Having the $H$-tensors at our disposal, it is now easy to systematically evaluate the perturbed operator traces \eqref{tr:canform}. Expanding
\be
\cO = \sum_{k = 0}^n M^{\a_1 \ldots \a_{2k}} \, D_{(\a_1} \cdots D_{\a_{2k})} \, 
\ee
with totally symmetric matrices $M^{\a_1 \ldots \a_{2k}}$, we have
\be
\begin{split}\label{trace:inv}
\Tr_i \left[ W(\Delta) \cO \right] = & \, \int_0^\infty ds \; \tilde{W}(s) \Tr_i \left[ \, \e^{-s \Delta} \, \sum_{k = 0}^n M^{\a_1 \ldots \a_{2k}} \, D_{(\a_1} \cdots D_{\a_{2k})}  \, \right] \\
= & \,
\int d^dx \sqrt{g} \, \int_0^\infty ds \; \tilde{W}(s) \, \tr_i \left[ \sum_{k = 0}^n \, M^{\a_1 \ldots \a_{2k}} \, H_{\a_1 \ldots \a_{2k}} \, \right] \\
= & \,
\frac{1}{(4\pi)^{d/2}} \, \int_0^\infty ds \, \tilde{W}(s) \, \sum_{k,l} \, s^k \, I_l \, . \\
\end{split}
\ee
Here, $I_l$ indicates the interaction monomials contained in the truncation ansatz, i.e., $I_2 = \int d^dx \sqrt{g} R, I_3 = \int d^dx \sqrt{g} R^2$, etc.

The final form of the result is most conveniently written in terms of the Mellin-transforms
\be
Q_n[W] = \int^\infty_0 ds \, s^{-n} \, \tilde{W}(s) \, ,
\ee
which we can re-express in terms of the original function $W$ as
\be\label{Qdef}
\begin{split}
Q_n[W] = & \frac{1}{\Gamma(n)} \int_0^\infty dz \, z^{n-1} \, W(z) \, , \qquad
Q_0[W] =  W(0) \, .
\end{split}
\ee
In view of practical computations, we furthermore note that there is a degeneracy between $Q$-functionals if their 
argument contains powers of the Laplacian as a factor
\be\label{Qrel}
Q_n[\Delta^p W] = \frac{\Gamma[n+p]}{\Gamma[n]} Q_{n+p}[W] \, , \quad n+p > 0 \, .
\ee

\smallskip

For a given truncation, the r.h.s.\ of the flow equation \eqref{FRGE} is then obtained by adding the relevant contributions of all the individual traces \eqref{trace:inv} and \eqref{trace:kin}. 
The desired $\beta$-functions governing the scale-dependence of the coupling constants are then read off as the coefficients multiplying the interaction monomials spanning a given truncation.
  
We close this section by highlighting the central virtues of the algorithm outlined above. Firstly, it is completely algebraic. No numerical integrations are needed in order to determine the $\beta$-functions. Secondly, the shape of the cutoff function $\cR_k$ is left arbritray. Thirdly, and most importantly, at no point in its construction does it make any reference to a particular choice of background. This is one of its main improvements compared to other computations. All previous explorations of the gravitational theory space, for instance, relied heavily on the fact that the background metric satisfies certain properties, as e.g., being the maximally symmetric metric on the $d$-sphere, or an Einstein metric. For the systematic construction of the derivative expansion in gravity theories, relaxing this technical limitation is of central importance.

%-------------------------------------------------------------
\section{Example: The Einstein-Hilbert truncation}
\label{sect:example}
%-------------------------------------------------------------
We now illustrate the working of the universal RG machine within  the simplest gravitational setting, the single-metric Einstein-Hilbert truncation in $d=4$ \cite{Reuter:1996cp,Reuter:2001ag,Lauscher:2001ya,Litim:2003vp,Codello:2008vh}. In this case, the expansion scheme is controlled by the curvature tensor of the background metric and the truncation retains all terms up to linear order in $\Rb$. In particular, this implies that terms which contain a covariant derivative of a curvature scalar (e.g. $\Db_\s \Rb_{\a\b\m\n}$, etc.) do not contribute to the truncation and can safely be neglected. Besides demonstrating that the new algorithm leads to reliable results, the main emphasis of this section is on demonstrating the background-independence of the resulting $\beta$-functions.\footnote{See also \cite{Codello:2008vh} for a related discussion.}
Instead of working with a specific background metric, our construction will leave $\gb_{\m\n}$ unspecified and the only technical assumption is that $\gb_{\m\n}$ is a (Euclidean) metric on a compact, closed, and complete manifold, which guarantees that the minimal TT-decomposition and heat-kernel expansion are well defined.

Splitting the averaged metric according to $g_{\m\n} = \gb_{\m\n} + h_{\m\n}$, the Einstein-Hilbert truncation corresponds to
 the following ansatz for the effective average action 
\be\label{EH:ansatz}
\Gamma_k[h, C, \bar{C}; \gb] = \Gamma_k^{\rm grav}[\gb+h] + S^{\rm gf}[h; \gb] + S^{\rm ghost}[h, C, \bar{C}; g] + S^{\rm aux} \, ,
\ee
where the gravitational part $\Gamma_k^{\rm grav}[g]$ is
\be\label{ansatzcan}
\Gamma_k^{\rm grav}[g] = \frac{1}{16 \pi G_k}  \int d^4x \sqrt{g} \, (2 \Lambda_k - R)  \, ,
\ee
and the gauge fixing and ghost actions are defined in \eqref{S.gf} and \eqref{Sghost} respectively (see Appendix \ref{App:E} for the derivation of these truncation-independent terms).
For convenience, we combine the scale-dependent dimensionful Newton's constant $G_k$ and cosmological constant $\Lambda_k$ into
\be\label{ucoup}
u_0 = \frac{\Lambda_k}{8 \pi G_k} \, , \quad u_1 = - \frac{1}{16 \pi G_k} \, ,
\ee
and introduce the dimensionless couplings
\be\label{gcoup}
g_0 = u_0 k^{-4} \, , \quad g_1 = u_1 k^{-2} \, .
\ee

Substituting the ansatz \eqref{EH:ansatz} and evoking the Landau-gauge decoupling theorem of App.\ \ref{sect:dec}, the FRGE \eqref{FRGE} reduces to the simple form
\be\label{trflow}
\p_t \Gamma_k = \cS^{\rm grav} + \cS^{\rm uni} \, .
\ee
Here, the $\cS^{\rm grav}$-trace captures the contributions of the physical degrees of freedom in the gravitational sector, the transverse-traceless tensor $h_{\m\n}^{\rm T}$ and the trace $h$, while
$\cS^{\rm uni}$ contains all terms that reoccur in any action for single metric gravity, including gauge fixing, ghost terms and Jacobians for the implemented decompositions. The latter are universal, in the sense that they do not dependent on the gravitational part of the effective average action \eqref{ansatzcan}. Their contribution is found in App.\ \ref{App:E} and reads
\be \label{tr:uni}
\cS^{\rm uni} = \tfrac{1}{(4\pi)^2} \int d^4x \sqrt{g}\; \Big\{
-2 Q_2[f^1] + R \left(
- \tfrac{5}{24} Q_1[f^1] - \tfrac{9}{8} Q_2[f^2] + Q_3[f^3] + 6 Q_4[f^4]
\right) \Big\} \, ,
\ee 
with the functions $f^n$ defined in eq.\ \eqref{fdef}.

%-------------------------------------------------------------
\subsection{The contribution of $\cS^{\rm grav}$}
%-------------------------------------------------------------
The contribution of the gravitational sector $\cS^{\rm grav}$ follows from first expanding \eqref{ansatzcan} to second order in the fluctuation field, and subsequently carrying out the  
TT-decomposition \eqref{mTTdec}. Retaining the terms built from the transverse-traceless tensor $h_{\m\n}^{\rm T}$ and the scalar trace $h$ only and dropping the bar from background quantities, this results 
in
\be \label{quad}
\Gamma_k^{\rm grav,quad} =
\frac{1}{2} \int d^4x \sqrt{g} \, \bigg\{
 h_{\alpha\beta}^{\rm T} \left[ \Kbb_{\rm 2T}^{\alpha\beta\m\n}
   + \Mbb_{\rm 2T}^{\alpha\beta\m\n} \right] \, h_{\m\n}^{\rm T}
 + h \left[ \Kbb_{0} + \Mbb_{0} \right] h
\bigg\} \, ,
\ee
with
\be \label{mathThT}
\begin{array}{lcl}
\mathbb{K}_{\rm 2T}^{\alpha\beta\m\n} & = & \, - g^{\a\m} g^{\b\n} \left(  u_1  \Dz  + u_0 \right) \, , \\
\mathbb{M}_{\rm 2T}^{\alpha\beta\m\n} & = & \, u_1( -g^{\a\m} g^{\b\n}   R + 2 R^{\a\m} g^{\b\n}  +2 R^{\a\m\b\n} ) \, , \\
\mathbb{K}_{0} & = & \,  \tfrac{3}{8} u_1 \Dz + \tfrac{1}{4} u_0 \, , \\
\mathbb{M}_{0} & = & \, 0 \, .
\end{array}
\ee
Notably, there are no cross-terms and a potential is only present in the transverse part.
The operators $\mathbb{K}_{\rm 2T}$ and $\mathbb{M}_{\rm 2T}$ should be understood as restricted to the transverse-traceless subspace. Instead of implementing this restriction explicitly, it is more convenient to leave them in the form \eqref{mathThT} and  employ the necessary projection operators \eqref{2Tproj} inside the trace later on.
It is precisely these projectors which lead to non-trivial terms that allow us to demonstrate the virtues of our algorithm.

%-------------------------------------------------------------
\subsubsection{The IR-cutoff}
%-------------------------------------------------------------
We now construct the matrix-valued IR cutoff $\cR_k$ for the gravitational sector, which is added to the action in the form
\be \label{S:cutoff}
\Delta_kS = \frac{1}{2} \int d^4x \sqrt{g} \left[ h^{\rm T}_{\a\b} \, \cR_{k,{\rm 2T}}^{\a\b\m\n} \, h^{\rm T}_{\m\n} + h \, \cR_{k,0} \, h \right] \, .
\ee
According to the Type Ic scheme \eqref{TypeIb}, we only regulate the kinetic terms $\mathbb{K}_i$, so that there are no curvature terms entering $\cR_k$. 
In the case at hand, this is achieved by 
\be
\begin{split}
\cR_{k,{\rm 2T}}^{\a\b\m\n} = & \, - u_1 g^{\a\m} g^{\b\n} \, R_k \, , \qquad
\cR_{k,0} = \, \tfrac{3}{8}  u_1 R_k \, .
\end{split}
\ee
The inclusion of $\cR_k$ then results in replacing
\be
\mathbb{K}_{\rm 2T} \rightarrow \mathbb{P}_{\rm 2T} = \mathbb{K}_{\rm 2T} + \cR_{k,{\rm 2T}} \, , \qquad
\mathbb{K}_{0} \rightarrow \mathbb{P}_{0} = \mathbb{K}_{0} + \cR_{k,0} \, ,
\ee
which, instead of $\Dz$, now include the regulated $P_k$ as arguments.

%-------------------------------------------------------------
\subsubsection{The perturbative inversion of $\Gamma_k^{(2)} + \cR_k$}
%-------------------------------------------------------------
Since the truncation ansatz \eqref{ansatzcan} contains only terms to first order in the curvature, it is sufficient to carry out the inversion neglecting all terms of the order $\mathbb{M}^2$. 
Adapting the general formulas \eqref{exp1} to independently invert the tensor and the scalar parts, we have
\be \label{2Tinverse}
\left[ \Gamma_k^{(2)} + \cR_k\right]_{h^{\rm T}h^{\rm T}}^{-1} = \,
\Pi_{\rm 2T} \cdot \left[ \frac{1}{\mathbb{P}_{\rm 2T}}
- \frac{1}{\mathbb{P}_{\rm 2T}} \Pi_{\rm 2T} \cdot \mathbb{M}_{\rm 2T} \cdot \Pi_{\rm 2T}  \frac{1}{\mathbb{P}_{\rm 2T}} \right] \cdot \Pi_{\rm 2T}
\ee
and
\be \label{Sinverse}
\left[ \Gamma_k^{(2)} + \cR_k\right]_{hh}^{-1} = \,
\Pi_{\rm tr}  \frac{1}{\mathbb{P}_{hh}} \Pi_{\rm tr} \, .
\ee
The projection operators $\Pi_{s}$ are defined in \eqref{2Tproj} and ensure that all terms are restricted to their proper subspaces.

%-------------------------------------------------------------
\subsubsection{Evaluation of the operator traces}
%-------------------------------------------------------------
In the following, let us denote the trace contributions by $\cT_{\rm s}^i$, where $s$ is the spin of the trace argument and $i$ enumerates the contributions. In the scalar sector, there is only the operator trace,
\be
\cT_0^1 = \frac{1}{2} \Tr_0 \left[ \, \frac{1}{P_{hh}} \, \p_t R_k^{\rm 0} \, \right]\,,
\ee
while in the tensorial sector we have
\be
\begin{split}
\cT_{\rm 2T}^1 = & \, \frac{1}{2} \Tr_2 \left[ \, \frac{1}{P_{\rm 2T}} \, \p_t R_k^{\rm 2T} \, \Pi_{\rm 2T} \, \right] 
\equiv  \, \frac{1}{2} \Tr_{\rm 2T} \left[ \frac{1}{P_{\rm 2T}} \, \p_t R_k^{\rm \rm 2T}  \right] \, , \\
\cT_{\rm 2T}^2 = & \, - \frac{1}{2} \Tr_2 \left[ \, \frac{1}{P_{\rm 2T}} \, \Pi_{\rm 2T} \cdot  \mathbb{M}_{\rm 2T} \cdot \Pi_{\rm 2T} \, \frac{1}{P_{\rm 2T}} \, \p_t R_k^{\rm \rm 2T} \, \Pi_{\rm 2T} \, \right] \, .
\end{split}
\ee
Here, $P_{\rm 2T}$ and $P_{0}$ are the {\it scalar parts} of the matrix valued operators $\mathbb{P}_{\rm 2T}$ and $\mathbb{P}_{0}$
\be
P_{\rm 2T} = - u_1 \, P_k - u_0 \, , \qquad P_{0} =  \tfrac{3}{8} \, u_1 \, P_k +\tfrac{1}{4}  u_0 \, .
\ee
The absence of contributions from commutators involving the projection operators is ensured by \eqref{traceless}.

The scalar and the first 2T-trace are easily found with the heat-kernel coefficients calculated in App.\ \ref{App:B3} and summarized in Table \ref{t.heatcoeff}. The $\cT_{\rm 2T}^2$-trace containing projector insertions is more complicated, but for the present truncation it is sufficient to simplify it to
\be\label{cT2T}
\cT_{\rm 2T}^2 = \, - \frac{1}{2} \Tr_2 \left[ \, \frac{1}{P_{\rm 2T}^2} \, \, \p_t R_k^{\rm \rm 2T} \, \mathbb{M}_{\rm 2T} \cdot \Pi_{\rm 2T}^{0r} \, \right] \, + \cO(R^2) \, ,
\ee
replacing the projector with its zero-order part \eqref{PiLzero} and commuting all covariant derivatives freely, since $\mathbb{M}_{\rm 2T}$ already contains one power of the curvature.
Its explicit evaluation requires the off-diagonal heat-kernel scheme explained in Subsection \ref{sect:nonmin} and represents a non-trivial demonstration of the proposed automatization technique. Using this scheme, the traces $\cT^1_0$, $\cT^1_{\rm 2T}$ and $\cT^2_{\rm 2T}$ evaluate to
\be
\begin{split}
\cT^1_0 = & \frac{1}{2} \,\frac{1}{(4\pi)^2} \, \int d^4x \sqrt{g}\; \Big\{
  Q_2[f^1_0] + \tfrac{1}{6} R Q_1[f^1_0]
\Big\}  \, , \\
\cT^1_{\rm 2T} = & \frac{1}{2} \,\frac{1}{(4\pi)^2} \, \int d^4x \sqrt{g}\; \Big\{
  5 Q_2[f^1_{\rm 2T}] - \tfrac{5}{6} \, R \, Q_1[f^1_{\rm 2T}]
\Big\} \, , \\
\cT^2_{\rm 2T} = & -\frac{1}{2} \frac{u_1}{(4\pi)^2}  \int d^4x \sqrt{g}\; \Big\{
- \tfrac{10}{3} \, R \, Q_2[f^2_{\rm 2T}]
\Big\} \, ,
\end{split}
\ee
with the dependence of the operator traces on the regularized Laplacian $P_k$ being captured by
\be \label{f2def}
\begin{split}
f^n_{\rm 2T} := & \frac{1}{(P_{\rm 2T})^n} \p_t R_{k}^{\rm 2T} \, , \qquad
f^n_0 :=  \frac{1}{(P_{0})^{n}} \p_t R_{k,0} \, .
\end{split}
\ee 
Combining these results, the full gravitational sector contribution then reads
\be\label{tr:grav}
\begin{split}
\cS^{\rm grav} =
\frac{1}{2} \,\frac{1}{(4\pi)^2} \, \int d^4x \sqrt{g} & \, \Big[ ( Q_2[f^1_0] + 5 Q_2[f^1_{\rm 2T}]) \\ & \;
+ \left( \tfrac{1}{6} Q_1[f^1_0] - \tfrac{5}{6} Q_1[f^1_{\rm 2T}] + \tfrac{10}{3} u_1 Q_2[f^2_{\rm 2T}] \right) R \Big] \, .
\end{split}
\ee
Notably, this is exactly the result obtained by earlier computations specifying the background metric to be the one on the four-sphere \cite{Machado:2007ea}.
We stress, however, that in our derivation of \eqref{tr:grav} the background metric $\gb_{\m\n}$ is left completely unspecified.
 This demonstrates the background-independence of the proposed algorithm and highlights its generalized applicability.

%-------------------------------------------------------------
\subsection{$\beta$-functions and non-Gaussian fixed point}
%-------------------------------------------------------------
The $\beta$-functions governing the running of Newton's constant and the cosmological constant can be read off
by substituting (\ref{tr:grav}) and (\ref{tr:uni}) into \eqref{trflow} and comparing the coefficients of $I_0 = \int d^4x \sqrt{g}$ and $I_1= \int d^4x \sqrt{g} R$.
The result is conveniently written in terms of the threshold functions
\be
\begin{split}
\Phi^p_n(\om) = & \frac{1}{\G(n)}\int_0^\infty dz\; z^{n-1} \frac{R^{(0)}(z)-zR^{(0)\prime}(z)}{(z+R^{(0)}(z)+\om)^p} \, , \\
\Pt^p_n(\om) = & \frac{1}{\G(n)}\int_0^\infty dz\; z^{n-1} \frac{R^{(0)}(z)}{(z+R^{(0)}(z)+\om)^p} \, ,
\end{split}
\ee
with $R_k(p^2) = k^2 R^{(0)}(p^2/k^2)$. Their relation to the $Q_n[f]$ is given by
\be
Q_n \left[\frac{1}{u_1(P_k+\om)^p}\p_t(u_1 R_k) \right] = 2k^{2(n-p+1)}\Big(
\Phi^p_n(\om/k^2) +\half (\tfrac{\p_t g_1}{g_1}+2) \Pt^p_n(\om/k^2) \Big) \, ,
\ee
which for the functions (\ref{f2def}) and (\ref{fdef}) yields
\be
\begin{split}
Q_n[f^p_{\rm 2T}] = & \; 2 (-g_1)^{1-p} k^{2(n-2p+2)} \Big(
\Phi^p_n(\tfrac{g_0}{g_1}) +\half (\tfrac{\p_t g_1}{g_1}+2) \Pt^p_n(\tfrac{g_0}{g_1}) \Big) \, , \\
Q_n[f^p_0] = & \; 2 (\tfrac{3}{8} g_1)^{1-p} k^{2(n-2p+2)}\Big(
\Phi^p_n(\tfrac{2}{3} \tfrac{g_0}{g_1}) +\half (\tfrac{\p_t g_1}{g_1}+2) \Pt^p_n(\tfrac{2}{3} \tfrac{g_0}{g_1}) \Big) \, , \\
Q_n[f^p] = & \; 2 k^{2(n-p+1)} \Phi^p_n(0) \, .
\end{split}
\ee

Inserting these relations into (\ref{tr:grav}) and (\ref{tr:uni}) and matching the power of the background curvature, the RG equation leads to
\be
\begin{split}
\p_t g_0 = & -4 g_0 +\tfrac{1}{(4\pi)^2} A_0(\tfrac{g_0}{g_1}) + \tfrac{1}{(4\pi)^2} \left( 1+\half\tfrac{\p_t g_1}{g_1} \right) B_0(\tfrac{g_0}{g_1}) \, , \\
\p_t g_1 = & -2 g_1 +\tfrac{1}{(4\pi)^2} A_1(\tfrac{g_0}{g_1}) + \tfrac{1}{(4\pi)^2} \left( 1+\half\tfrac{\p_t g_1}{g_1} \right) B_1(\tfrac{g_0}{g_1}) \, ,
\end{split}
\ee
which can be solved for the $\beta$-functions for the dimensionless Newton's and cosmological constant ($g$ and $\l$) with the use of the definitions (\ref{ucoup}) and (\ref{gcoup}). This yields
\be\label{betafct}
\begin{split}
\p_t \l = & (\eta_N - 2)\l + \frac{g}{2\pi} \left( A_0(\l) - \half\eta_N B_0(\l) \right) \, , \\
\p_t g = & (\eta_N + 2)g \, ,
\end{split}
\ee
with the anomalous dimension of Newtons constant
\be
\eta_N =  \frac{2 g A_1(\l)}{2\pi + g B_1(\l)} \, , 
\ee
and the cutoff shape dependent functions
\be
\begin{split}
A_0(\l) = & 5 \Phi_2^1(-2\l) + \Phi_2^1(-\tfrac{4}{3}\l) - 4 \Phi_2^1(0) \, , \\
B_0(\l) = & 5 \Pt_2^1(-2\l) + \Pt_2^1(-\tfrac{4}{3}\l) \, , \\
A_1(\l) = & -\tfrac{5}{6} \Phi_1^1(-2\l) + \tfrac{1}{6}\Phi_1^1(-\tfrac{4}{3}\l) - \tfrac{10}{3} \Phi_2^2(-2\l) \\
& -\tfrac{5}{12} \Phi_1^1(0) - \tfrac{9}{4} \Phi_2^2(0) + 2 \Phi_3^3(0) + 12 \Phi_4^4(0) \, , \\
B_1(\l) = & -\tfrac{5}{6} \Pt_1^1(-2\l) + \tfrac{1}{6}\Pt_1^1(-\tfrac{4}{3}\l) - \tfrac{10}{3} \Pt_2^2(-2\l) \, .
\end{split}
\ee

Evaluated with the Optimized Cutoff $R^{(0)}(z)=(1-z)\Theta(1-z)$ \cite{Litim:2001up}, this system has, in addition to the Gaussian fixed point, a non-Gaussian fixed point (NGFP), in agreement with previous works \cite{Reuter:2001ag,Lauscher:2001ya,Litim:2003vp,Codello:2008vh}. This NGFP is located at
\be\label{NGFP}
g_* = 1.0021 \, , \hspace{1cm} \l_* = 0.134414 \, , \hspace{1cm} g_* \l_* = 0.134696 \, ,
\ee
and has the stability coefficients
\be\label{NGFPstab}
\theta = 2.37141 \pm i\, 2.27954 \, .
\ee

The $\beta$-functions \eqref{betafct} and numerical values for the NGFP \eqref{NGFP} and \eqref{NGFPstab} are very close, but not identical to the ones obtained in \cite{Codello:2007bd,Machado:2007ea}. This difference can be traced back to the use of the two different cutoff schemes (Type I vs.\ Type Ic), which slightly alters the contribution steming from the gauge-fixing part of the action. We find it very encouraging, though, that both cutoffs lead to the same qualitative results.

%-------------------------------------------------------------
\section{Discussion and future perspectives}
\label{sect:disc}
%-------------------------------------------------------------
In this paper, we analyzed the structure of Wetterich-Type Functional Renormalization Group Equations \eqref{FRGE}, which provide an exact description of the Wilsonian renormalization group flow of a given theory. The flow equation contains the same physical information as the path-integral formulation, but is generally much more manageable. The prime computational obstacle when extracting non-perturbative physics from this equation is the evaluation of the operator traces appearing on its right-hand side.

As the central result of this paper we presented an explicit algorithm, the universal RG machine, that bypasses this obstacle
by relating the operator traces to the off-diagonal Heat-Kernel expansion of a standard Laplace-Operator \cite{DeWitt:1975ys,Christensen:1978yd} (for reviews and further references, see also \cite{Decanini:2005eg,Anselmi:2007eq}). The explicit steps in the algorithm are detailed at the beginning of Sect.\ \ref{sect:algo}
and allow for the construction of the $\beta$-functions arsing from a given truncation by purely algebraic methods. Each of these steps can be handled by computer algebra software, which is highly convenient for the treatment of truncations beyond a certain complexity, where the $\beta$-functions arise from the sum of a large number of terms. As a welcome byproduct, the construction lifts the technical restriction of working with a particular background, manifestly showing the background independence of the formalism. Thus, the algorithm provides access to information that is otherwise restricted by the choice of the background.

These features open the door to a multitude of applications which up to now where out of computational reach. Furthermore, while our motivation for constructing the universal RG machine originates from studying the renormalization group flow of gravity in order to obtain further insights into the asymptotic safety conjecture, we stress that the applications of this construction are not limited to gravity and the algorithm is easily adapted to address open questions within other gauge theories.

As an illustration of our algorithm, we re-derived the $\beta$-functions of the single-metric Einstein-Hilbert truncation \cite{Reuter:2001ag,Lauscher:2001ya,Litim:2003vp,Codello:2008vh}. The resulting flow equation was evaluated for a generic background metric, demonstrating that the resulting $\beta$-functions depend on the gauge-fixing and cutoff-scheme, but are explicitly independent of the choice of background metric.
Moreover, the new type of cutoff, intrinsic to the universal RG machine, gives rise to the same fixed-point structure found within all other cutoff-schemes \cite{Codello:2008vh,Reuter:2007rv}.

The construction of the $\beta$-functions is considerably simplified by working in the geometric Landau gauge \eqref{gaugechoice}. As demonstrated in App.\ \ref{sect:dec}, this choice leads to a factorization of the operator traces into a gravitational sector, containing only the contribution of the transverse-traceless and trace parts of the metric fluctuations, and a ``universal sector" which captures the contributions of the gauge fixing, ghost, and auxiliary terms arising from the applied field decompositions. Once computed to sufficient order, the latter sector can be recycled for any truncation of the gravitational effective average action, as it does not depend on the choice of the gravitational part of $\Gamma_k$.

A basic building block of our algorithm are the heat-kernel coefficients for transverse vectors and transverse-traceless matrices, which we computed on a manifold without boundary up to second order in the curvature. The result is given in Table \ref{t.heatcoeff}. While independent of the choice of metric, the result depends on the topology of the manifold, which affects the computation via the zero modes of the operators implementing the transverse decomposition of a vector field. Based on this observation, it is natural to expect that this topology-dependence will carry over to the gravitational $\beta$-functions. Thus, we expect that at a sufficient level of sophistication (which is beyond the Einstein-Hilbert case analyzed in Sect.\ \ref{sect:example}) the gravitational $\beta$-functions will be sensitive to topological effects.\footnote{A similar observation has already been made in the context of conformally reduced gravity \cite{Reuter:2008qx}, where the computations on a spherical and flat Minkowski space led to slightly different $\beta$-functions.}

Notably, the universal RG machine aids practical computations in any expansion scheme of the flow equation, such as a vertex expansion in a suitable background field.
This has already been demonstrated in the context of computing the ghost-wave-function renormalization for gravity \cite{Groh:2010ta}, 
where the expansion is controlled by the background ghost fields \cite{Eichhorn:2009ah,Groh:2010ta,Eichhorn:2010tb}. In a similar spirit, one can systematically carry out bimetric computations along the lines \cite{Manrique:2010mq, Manrique:2010am}, providing a handle on, e.g., the running of the gauge-parameters. 

In the case of gravity, the presented algorithm allows us to extend the explored theory space into hitherto uncharted terrain, facilitating the systematic derivative expansion of the gravitational effective average action \cite{Reuter:1996cp}. This task is closely linked to determining the fixed point structure and number of relevant couplings of the theory. Using our results, it is now straightforward to complete the study of four-derivative truncations initiated in \cite{Lauscher:2002sq,Benedetti:2009rx}, by including the scale-dependence of the Gauss-Bonnet term. This issue will be addressed in a forthcoming publication. Lastly, we note that, though requiring an improved understanding of the off-diagonal heat-kernel expansion going beyond the results presented here, the inclusion of the Goroff-Sagnotti term \cite{Goroff:1985sz,vandeVen:1991gw} can also be in principle implemented along the same lines.

%-------------------------------------------------------------
\section*{Acknowledgments}
%-------------------------------------------------------------
%
We are greatful to A.\ Codello, V.\ Cort\'es, R.\ Percacci, and M.\ Reuter for helpful discussions.
P.F.M.\ and F.S.\ thank the Perimeter Institute for hospitality and financial support
during the initial stages of the project. P.F.M.\ further thanks the Institute of Theoretical Physics of Utrecht University and R. Loll for their support in the early stages of this work.
The research of K.G.\ and F.S.\ is supported by the Deutsche Forschungsgemeinschaft (DFG)
within the Emmy-Noether program (Grant SA/1975 1-1).

%-------------------------------------------------------------
\begin{appendix}
%-------------------------------------------------------------
%-------------------------------------------------------------
\section{Commutator relations}
\label{sect:com}
%-------------------------------------------------------------
Commutator relations between covariant derivatives are an important building block of the universal RG machine. In this section, we summarize the most important commutator relations for the case of gravity, which also occur in the example given in Sect.\ \ref{sect:example}.

The commutator of two covariant derivatives acting on an arbitrary tensor $\phi_{\a_1 \a_2 \ldots \a_n}$ is given by
\be\label{basiscom}
\left[ D_\m \, , D_\n \right] \phi_{\a_1 \a_2 \ldots \a_n} = \sum_{k=1}^n \, R_{\m\n \a_k}{}^\rho \, \phi_{\a_1 \ldots \a_{k-1} \rho \a_{k+1} \ldots \a_n} \, .
\ee
Based on this, it is convenient to have commutators involving the Laplacian $\D=-D^\m D_\m$ acting on scalars $\phi$, vectors $\phi_\a$ and symmetric matrices $\phi_{\a\b}$
\be\label{exactcom}
\begin{array}{ll}
 \left[ \, D_\mu \, , \, \D \, \right] \phi  & =  R_\m{}^\a D_\a \phi \, , \\
 \left[ \, D_\mu \, , \, \D \, \right] \phi_\rho  &  =  R_\m{}^\a D_\a \phi_\rho - 2 R^\a{}_\m{}^\b{}_\rho \, D_\a \phi_\b  - (D_\a  R^\a{}_\m{}^\b{}_\rho ) \phi_\b \, , \\
 \left[ \, D_\mu \, , \, \D \, \right] \phi_{\r\s}  &  = R_\m{}^\a D_\a \phi_{\r\s}
- 4 R_{\a\m}{}^\b{}_{(\r} \, D^\a \phi_{\s)\b} - 2 \left( D_\a R^\a{}_\m{}^\b{}_{(\r} \right) \phi_{\s)\b} \, .
\end{array}
\ee
For later reference we also note the double-commutators
\be\label{doublecom}
\begin{array}{ll}
\left[ \left[ D_\m, \D \right], \D \right] \phi & =  R_\m{}^\a R_\a{}^\b D_\b \phi + \cO(DR) \, , \\
\left[ \left[ D^\gamma, \D \right], \D \right] \phi_{\gamma\b} & =
- R^{\gamma\n} \left[ R_\n{}^\m D_\m \phi_{\gamma \beta} + 4 R^\m{}_{(\gamma \n\a} D^\a \phi_{\beta)\m} \right] \\
& \quad + 2 R^\lambda{}_\b{}^{\gamma \n} \left[ R_\n{}^\m D_\m \phi_{\gamma\lambda} + 4 R^\m{}_{(\gamma \n\a} D^\a \phi_{\lambda)\m} \right] + \cO(DR)\, .
\end{array}
\ee

These basic commutator relations can be used to expand the commutator of a covariant derivative with an arbitrary function of the Laplacian in terms of the curvature. As a prerequisite, one verifies
\be\label{comexp}
\left[ D_\m \, , \, \e^{-s\D} \right] \, \phi_{\a_1 \ldots \a_n} = \e^{-s\D} \, \left\{ -s \left[ D_\m,\D \right] + \frac{s^2}{2} \left[ \left[ D_\m, \D \right], \D \right] \right\} \, \phi_{\a_1  \ldots \a_n} + \cO(DR, R^3) \, .
\ee
It is straightforward to extend \eqref{comexp} to functions of the Laplacian,
\be\label{gencom}
\left[ \, D_\mu \, , \, f(\Delta) \, \right] \phi_{\a_1 \ldots \a_n} = \Big\{ f^\prime(\Delta) \left[ \, D_\mu \, , \, \D \, \right] +\half f^{\prime\prime}(\Delta) \left[ \left[ D_\m, \D \right], \D \right] \Big\} \phi_{\a_1  \ldots \a_n} + \cO(DR, R^3) \, .
\ee
Here the prime denotes the derivative of $f(x)$ with respect to its argument. This result can be established by replacing $f(\Delta)$ by its Laplace-transform, subsequently substituting \eqref{comexp} and transforming back.

Finally, for practical computations, it is useful to have explicit expressions for the commutators of the projection operators \eqref{1Tproj} and \eqref{2Tproj}. For transverse vectors, one can use eqs.\ \eqref{cantrace}, \eqref{basiscom}, and \eqref{gencom} to show that
\be\label{1Tprojid}
\begin{split}
\left[ \Pi_{\rm T} \right]_\m{}^\b \, f(\Delta) \, \left[ \Pi_{\rm T} \right]_\b{}^\n \, \phi_\n 
= & \, \left[ \Pi_{\rm T} \right]_\m{}^\n \,  f(\Delta) \, \, \phi_\n 
-  \, f^\prime(\Delta) \,  \left[C_{\rm T} \right]_\m{}^\n \, \phi_\n \, ,
\end{split}
\ee
where
\be\label{1Tprojcom}
\left[C_{\rm T} \right]_\m{}^\n = \left( \tfrac{1}{\Delta} R_\m{}^\a D_\a + \tfrac{1}{\Delta^2} R^{\a\b} D_\m D_\a D_\b \right) \, D^\n \, , 
\ee
and all expressions are valid up to terms $\cO(R^2, D R)$.
Analogously, one can show that
\be\label{2Tprojid}
\begin{split}
\left[ \Pi_{\rm 2T} \right]_{\m\n}{}^{\r\s} \, f(\Delta) \, \left[ \Pi_{\rm 2T} \right]_{\r\s}{}^{\a\b} \, \phi_{\a\b} 
= & \, \left[ \Pi_{\rm 2T} \right]_{\m\n}{}^{\a\b} \,  f(\Delta) \, \, \phi_{\a\b}
- \, f^\prime(\Delta) \,  \left[C_{\rm 2T} \right]_{\m\n}{}^{\a\b} \, \phi_{\a\b} \, , 
\end{split}
\ee
with the corresponding commutator
\be\label{2Tprojcom}
\begin{split}
\left[C_{\rm 2T} \right]_{\m\n}{}^{\a\b} = & \, \left[ \Pi_{\rm 2T} \right]_{\m\n}{}^{\r\s} 
\left( -2 R_{(\r}{}^\l \delta_{\s)}^{(\a} D^{\b)} D_\l + 4 R^\l{}_{(\r}{}^\tau{}_{\s)} D_\l \delta_\tau^{(\a} D^{\b)}  \right) \, \tfrac{1}{\D} \,  .
\end{split}
\ee
In order to arrive at this expression, it is useful to observe that the bracket is generated by the first term in \eqref{PiLzero}. All other commutators are either proportional to $g_{\a\b}$, $D_\a$, or $D_\b$ and are therefore annihilated when contracted with $\Pi_{\rm 2T}$.

A central property of the tensors \eqref{1Tprojcom} and \eqref{2Tprojcom} is that they are traceless with respect to the internal indices
\be\label{traceless}
\begin{split}
& \tr_1 \left[C_{\rm T} \right]_\m{}^\n  = \delta^\m_\n \left[C_{\rm T} \right]_\m{}^\n = 0 \, , \\
& \tr_2 \left[ C_{\rm 2T} \right]_{\m\n}{}^{\a\b} = \half 
\left( \delta^\m_\a \delta^\n_\b + \delta^\m_\b \delta^\n_\a \right) \left[ C_{\rm 2T} \right]_{\m\n}{}^{\a\b} = 0 \, .
\end{split}
\ee
This ensures that commutators of this type do not contribute to the Einstein-Hilbert truncation.

%-------------------------------------------------------------
\section{Transverse-traceless decomposition}
\label{sect:ttdec}
%-------------------------------------------------------------
Another central ingreedient of the universal RG machine is the York-decomposition of the fluctuation fields into their transverse and longitudinal parts \cite{York:1973ia,YorkTT}. In order for the decomposition to be well-defined, we will assume that the background is a closed (i.e., compact without boundary) and complete manifold.\footnote{Strictly speaking, the assumption of the background being closed can be relaxed and one could consider asymptotically flat backgrounds when appropriate assumptions on the fall-off of the metric and fluctuation fields are implemented.} Under these assumptions the decomposition is unique, up to ambiguities associated with Killing vectors or conformal Killing vectors of the background which constitute zero modes of the projection operators implementing the decomposition. In the sequel, we will first discuss the decomposition of vector fields, before turning to symmetric tensors in Subsection \ref{sect:tttensor}.
%
%-------------------------------------------------------------
\subsection{Decomposition of vector fields}
\label{sect:ttvec}
%-------------------------------------------------------------
Let us start with the decomposition of a generic vector field, which typically appears in the gravitational gauge- and ghost-sector. The split into its transversal and longitudinal part follows from \eqref{vec:dec} and is subject to the differential constraint \eqref{vec:dconst}.
The systematic implementation of this decomposition at the level of the universal RG machine requires the introduction of suitable projection operators on the space of (unconstrained) vectors
\be\label{1Tproj}
[\unit_1]_\m{}^\n = \delta_\m{}^\n \, , \qquad [\Pi_{\rm L}]_\m{}^\n = - D_\m \Delta^{-1} D^\n \, , \qquad
[\Pi_{\rm T}]_\m{}^\n = [\unit_1]_\m{}^\n - [\Pi_{\rm L}]_\m{}^\n \, .
\ee
These satisfy the natural projector properties
\be
\Pi_{\rm L} \cdot \Pi_{\rm L} = \Pi_{\rm L} \, , \qquad \Pi_{\rm T} \cdot \Pi_{\rm T} = \Pi_{\rm T} \, , \qquad \Pi_{\rm L} \cdot \Pi_{\rm T} = \Pi_{\rm T} \cdot \Pi_{\rm L} = 0 \, ,
\ee
which also ensure the orthogonality of the decomposition. Applying these projectors to \eqref{vec:dec}, it is then straightforward to establish that
\be
\left[ \Pi_{\rm T} C \right]_\m = C^{\rm T}_\m \, , \qquad \left[ \Pi_{\rm L} C \right]_\m = D_\m \eta \, .
\ee

All of these relations are exact in the sense that they do not refer to a systematic expansion in the background curvature. For practical computations, however, it is convenient to have an explicit expression for $\Pi_{\rm L}$, where the Laplacian is to the very left, or very right of the matrix structure:
\be\label{1Tprojpert}
\begin{split}
\left[ \Pi_{\rm L} \right]_\m{}^\n C_\n = & \,
\left[ - D_\m D^\n \Delta^{-1} + D_\m D_\a R^{\a\n} \Delta^{-2} + \cO(R^2) \right] C_\n \\
= & \,
\left[ - \Delta^{-1} D_\m D^\n  + \Delta^{-2} R_{\m\a} D^\a D^\n + \cO(R^2) \right] C_\n \, .
\end{split}
\ee
These can be obtained as a perturbative series in $R$ by making use of the commutator relations \eqref{gencom}.

%-------------------------------------------------------------
\subsection{Decomposition of symmetric tensor fields}
\label{sect:tttensor}
%-------------------------------------------------------------
The gravitational fluctuations $h_{\m\n}$ can be decomposed in a transverse-traceless part $h_{\m\n}^{\rm T}$, a vector $\xi_\m$, and a scalar $h$ encoding the trace-part. The explicit deconstruction follows from \eqref{mTTdec} with the component fields satisfying \eqref{TTconst}. 
The ``standard'' TT-decomposition, employed for example in \cite{Dou:1997fg,Lauscher:2001ya}, and needed to eliminate all the $\Dbb$-terms, includes a splitting of the vector $\xi_\m$ into its longitudinal and transverse parts, as in \eqref{vec:dec}. The reason we are not implementing that here is that it represents a superfluous complication for the projection on the ${\rm 2T}$-component of the fluctuation, and thus also for the derivation of the associated heat kernel coefficients in App.~\ref{App:B3}. In order to distinguish the decomposition \eqref{mTTdec} from the ``standard'' TT-decomposition, we will refer to \eqref{mTTdec} as ``minimal'' TT-decomposition.

When evaluating operator traces on the space of ${\rm 2T}$-fields it is again convenient to introduce covariant projection operators onto the transverse-traceless, vector, and scalar subspaces. These can be constructed from
\be\label{2Tproj}
\begin{split}
[ \, \unit_2 \, ]_{\a\b}{}^{\rho\sigma} = & \, \half \left( \delta_\a{}^\rho \delta_\b{}^\sigma + \delta_\a{}^\sigma \delta_\b{}^\rho  \right) \, , \\
[ \, \Pi_{\rm tr} \, ]_{\a\b}{}^{\rho\sigma} = & \, \tfrac{1}{d} g_{\a\b} g^{\rho\sigma} \, , \\
\left[ \, \Pi_{\rm 2L} \,  \right]_{\a\b}{}^{\rho\sigma} = & \, \left[P_1\right]_{\a\b}^\m \, \left[ P_2^{-1} \right]_\m^\n \left(-D^\gamma\right) \left[ \, \unit_2 - \Pi_{\rm tr} \, \right]_{\gamma\nu}{}^{\r\s} \, . 
\end{split}
\ee
Here, $\unit_2$ is the unit-operator on the space of symmetric matrices and
\be\label{Pmat}
\left[P_1\right]_{\a\b}^\m = 2 D_{(\a} \delta_{\b)}^\m  - \tfrac{2}{d} g_{\a\b} D^\m \, , \quad
\left[ P_2^{-1} \right]_\m^\n = \left[ \Delta \delta^\m_\n - R^\m{}_\n - \tfrac{d-2}{d} D^\m D_\n \right]^{-1} \, .
\ee
Applying these projectors to \eqref{mTTdec}, it is easy to verify that they project $h_{\m\n}$ onto its vector- and trace-part respectively,
\be\label{proj2}
\begin{split}
\left[ \, \Pi_{\rm 2L} h \, \right]_{\m\n} = & [P_1]^{\a}_{\m\n} \xi_\a = D_\m \xi_\n + D_\n \xi_\m - \tfrac{2}{d} g_{\m\n} D^\alpha \xi_\alpha \, , \qquad
\left[ \, \Pi_{\rm tr} h \, \right]_{\m\n} = \tfrac{1}{d} \, g_{\mu\nu} \, h \, .
\end{split}
\ee
Furthermore, they satisfy the standard projector identities
\be
\Pi_{\rm tr} \cdot \Pi_{\rm tr} = \Pi_{\rm tr} \, , \qquad \Pi_{\rm 2L} \cdot \Pi_{\rm 2L} = \Pi_{\rm 2L} \, , \qquad \Pi_{\rm tr} \cdot \Pi_{\rm 2L} = \Pi_{\rm 2L} \cdot \Pi_{\rm tr} = 0 \, ,
\ee
ensuring the orthogonality of the decomposition on an arbitrary background. Based on the properties \eqref{proj2}, the projector on the transverse-traceless part $h^{\rm T}_{\m\n}$ is given by
\be\label{2Tprojector}
\Pi_{\rm 2T} = \unit_2 - \Pi_{\rm 2L} - \Pi_{\rm tr} \, , \qquad \left[ \Pi_{\rm 2T} h \right]_{\m\n} = h^{\rm T}_{\m\n} \, .
\ee
This last projector is a key ingredient for the construction of the heat-kernel expansion on the space of $h^{\rm T}_{\m\n}$-fields subject to the constraints \eqref{TTconst} in Sect.\ \ref{App:B3}.

The presence of the complicated pseudo-differential operator $\left[ P_2^{-1} \right]_\m^\n$ severely complicates working with the projector $\Pi_{\rm 2T}$. To bypass this obstruction, it is necessary to recast this operator into a power series in the background curvature,
\be\label{cres}
\begin{split}
\left[ P_2^{-1} \right]_\a^\b \xi_\b = & \bigg[ 
 \tfrac{1}{\D} \delta_\a{}^\b + \half \tfrac{d-2}{d-1} D_\a D^\b \tfrac{1}{\D^2} + R_\a{}^\b \tfrac{1}{\D^2} + \tfrac{d-2}{d-1} R_\a{}^\m D_\m D^\b \tfrac{1}{\D^3} \\
 & \qquad + \half \tfrac{(d-2)^2}{(d-1)^2} R^{\m\n} D_\m D_\n D_\a D^\b \tfrac{1}{\D^4}
\bigg] \xi_\b + \cO(R^2, D R) \, .
\end{split}
\ee
In order to derive this expansion, we abbreviate $q \equiv (d-2)/d < 1$ and read the formal inverse \eqref{Pmat} as a geometric series in the $q$- and curvature terms. The perturbative
inverse then assumes the form
\be \label{inverse-defP}
\left[ P_2^{-1} \right]_\a^\b \xi_\b = \left( \D^{-1} \delta_\a{}^\b  + \left[P_{2,0} \right]_\a^\b + \left[P_{2,R} \right]_\a^\b \right) \xi_\b+\cO(R^2, D R) \, ,
\ee
where $P_{2,0}$ and $P_{2,R} $ capture the resummed contribution of the $q$-term at zeroth and first order in $R_\a^\b$, respectively. Explicitly, we can write
\be \label{P_20}
\begin{split}
\left[ P_{2,0} \right]_\a^\b \xi_\b & = \sum_{n=0}^{\infty} \,  q^{n+1} \, \frac{1}{\D} D_\a \, \left(D^\g\frac{1}{\D} D_\g\right)^n \, D^\b \, \frac{1}{\D} \, \xi_\b \\
& =  \sum_{n=0}^{\infty} \, q^{n+1} \, \frac{1}{\D} \, D_\a \, \left(  -1 +  \frac{1}{\D^2} R^{\m\n} D_\m D_\n   \right)^n \,  D^\b \frac{1}{\D} \, \xi_\b \\
& = \left[ \frac{d-2}{2(d-1)} \, \frac{1}{\D} D_\a D^\b \frac{1}{\D} + \frac{(d-2)^2}{4 (d-1)^2} R^{\m\n} D_\m D_\n D_\a D^\b \frac{1}{\D^4} \right] \, \xi_\b \, ,
\end{split}
\ee
and
\be \label{P_2R}
\begin{split}
& \left[P_{2,R} \right]_\a^\b \xi_\b \\  & = \left[ \frac{1}{\D} R_\a{}^\b \frac{1}{\D} + \sum_{n=1}^{\infty} \frac{q^n}{\D^{n+2}} \left( 2 R_{(\a}{}^\m D_\m D^{\b)} D^{2(n-1)}   +(n-1) D^{2(n-2)} R^{\m\n} D_\m D_\n D_\a D^\b  \right) \right] \xi_\b \\
& = \left[ \frac{1}{\D} R_\a{}^\b \frac{1}{\D} + \sum_{n=1}^{\infty} (-q)^n \left( -\frac{2}{\D^3} R_{(\a}{}^\m D_\m D^{\b)}   +\frac{n-1}{\D^4}  R^{\m\n}D_\m D_\n D_\a D^\b  \right)
\right] \xi_\b \\
& = \left[ \frac{1}{\D} R_\a{}^\b \frac{1}{\D} + \frac{d-2}{d-1} \frac{1}{\D^3} R_{(\a}{}^\m D_\m D^{\b)}   + \frac{(d-2)^2}{4 (d-1)^2} \frac{1}{\D^4}  R^{\m\n} D_\m D_\n D_\a D^\b \right] \xi_\b \, .
\end{split}
\ee
Substituting them into \eqref{inverse-defP} and applying the commutator \eqref{gencom} to the first term in \eqref{P_20} finally proves \eqref{cres}. The resummations \eqref{P_20} and \eqref{P_2R} are exact up to terms $\cO(R^2, DR)$. Higher orders can be calculated in the same way, taking higher order commutators into account.

With the perturbative inverse of $P_2^{-1}$ at hand, it is now straightforward to give the projector $\Pi_{\rm 2L}$ in terms of a perturbative series in the background curvature. Omitting terms $\cO(R^2, DR)$, it is convenient to express the series as
\be\label{LLRpara}
\begin{split}
[\, \Pi_{\rm 2L} \, ]_{\rm \m\n}{}^{\r\s} \phi_{\r\s} \approx & \, \left[ \, \Pi_{\rm 2L}^{0l} + \Pi_{\rm 2L}^{1l} + \Pi_{\rm 2L}^{1} \right]_{ \m\n}{}^{\r\s} \phi_{\r\s} \\
= & \, \left[ \, \Pi_{\rm 2L}^{0r} + \Pi_{\rm 2L}^{1r} + \Pi_{\rm 2L}^{1} \right]_{\m\n}{}^{\r\s} \phi_{\r\s} \, ,
\end{split}
\ee
where, $\Pi_{\rm 2L}^{0l}$ and $\Pi_{\rm 2L}^{0r}$ denote the part with no curvature terms and the Laplacians written to the left and right, $\Pi_{\rm 2L}^{1l}$ and $\Pi_{\rm 2L}^{1r}$ capture the corresponding commutator contributions from moving the Laplacians, and $\Pi_{\rm 2L}^{1}$ is the linear curvature contribution originating from the inversion formula \eqref{cres}. A straightforward computation shows
\be\label{PiLzero}
\begin{split}
[\, \Pi_{\rm 2L}^{0r} \, ]_{\a\b}{}^{\r\s} = & \,
- 2 D_{(\a} \delta_{\b)}^{(\rho} D^{\s)} \tfrac{1}{\D} + \tfrac{1}{d-1} g^{\r\s} D_{(\a} D_{\b)} \tfrac{1}{\D} + \tfrac{1}{d-1} g_{\a\b} D^{(\r} D^{\s)} \tfrac{1}{\D} \\ & \, 
- \tfrac{d-2}{d-1} D_{(\a} D_{\b)} D^{(\r} D^{\s)} \tfrac{1}{\D^2} + \tfrac{1}{d(d-1)} g_{\a\b} g^{\r\s} \, , \\
[\, \Pi_{\rm 2L}^{0l} \, ]_{\a\b}{}^{\r\s} = & \,
- 2 \tfrac{1}{\D} D_{(\a} \delta_{\b)}^{(\rho} D^{\s)}  + \tfrac{1}{d-1} \tfrac{1}{\D} g^{\r\s} D_{(\a} D_{\b)}  + \tfrac{1}{d-1} \tfrac{1}{\D} g_{\a\b} D^{(\r} D^{\s)}  \\ & \, 
- \tfrac{d-2}{d-1} \tfrac{1}{\D^2} D_{(\a} D_{\b)} D^{(\r} D^{\s)}  + \tfrac{1}{d(d-1)} g_{\a\b} g^{\r\s} \, ,
\end{split}
\ee
with the resulting commutator pieces reading
\be\label{Pi1lr}
\begin{split}
[\, \Pi_{\rm 2L}^{1r} \, ]_{\a\b}{}^{\r\s} = & \, \left(
2 D_{(\a} \delta_{\b)}^\tau - \tfrac{1}{d-1} g_{\a\b} D^\tau + \tfrac{2(d-2)}{(d-1)} D_{(\a} D_{\b)} D^\tau \tfrac{1}{\D} \right) \times \\ & \qquad 
\times \left( R^{\l\m} D_\l \delta^\n_\tau - 2 R_\tau{}^{\m\l\n} D_\l \right)
 \left[ \, \unit_2 - \Pi_{\rm tr} \, \right]_{\m\n}{}^{\rho\sigma}
\tfrac{1}{\Dz^2} \\ & \,
- \tfrac{d-2}{d(d-1)} \, g_{\a\b} \, R^{\l\m} \, D_\l D^\n \, \tfrac{1}{\Dz^2}\, \left[ \, \unit_2 - \Pi_{\rm tr} \, \right]_{\m\n}{}^{\rho\sigma} \, , \\
[\, \Pi_{\rm 2L}^{1l} \, ]_{\a\b}{}^{\r\s} = & \, \ \bigg[ \tfrac{1}{\D^2} 
\left( 2 R_{(\a}{}^\l D_\l \delta_{\b)}^\n - 4 R^\m{}_{(\a}{}^\n{}_{\b)} D_\m  + \tfrac{2}{d(d-1)} g_{\a\b} R^{\m\n} D_\m \right) \\
&
\; \; + \tfrac{2(d-2)}{d-1} \tfrac{1}{\D^3} \left( 2 R_{(\a}{}^\l D_{\b)} D_\l D^\n - 2 R^\l{}_\a{}^\s{}_\b D_\l D_\s D^\n - D_{(\a} D_{\b)} D_\m R^{\m\n} \right)
\bigg] \times \\ & \; \; \times
\, D^\g \, \left[ \, \unit_2 - \Pi_{\rm tr} \, \right]_{\g\n}{}^{\rho\sigma} \, .
\end{split}
\ee
Finally, the commutator free piece appearing in both expressions is
\be\label{Pi1gen}
\begin{split}
[\, \Pi_{\rm 2L}^{1} \, ]_{\a\b}{}^{\r\s} = & \, 
\left( 2 D_{(\a} \delta_{\b)}^\m - \tfrac{2}{d} g_{\a\b} D^\m \right) \times \\
& \, \times \left( R_\m{}^\n \tfrac{1}{\D^2} + \tfrac{d-2}{d-1} R_\m{}^\g D_\g D^\n \tfrac{1}{\D^3} - \tfrac{(d-2)^2}{2(d-1)^2} R^{\g\d} D_\g D_\d D_\m D^\n \tfrac{1}{\D^4}  \right) \times \\
& \, \times \left( D^{(\rho} \delta^{\sigma)}_\n - \tfrac{1}{d} g^{\r\s} D_\n \right) \, .
\end{split}
\ee
Recall that, at the level of the Einstein-Hilbert truncation, all covariant derivatives in \eqref{Pi1lr} and \eqref{Pi1gen} can be commuted freely, as we drop all terms of order $\cO(R^2, D R)$.
The expansion \eqref{LLRpara} is the central ingredient in practical computations to handle the projection operators in a trace argument.

%-------------------------------------------------------------
\subsection{Jacobians for the decomposition of vector and symmetric tensor fields}
\label{sect:aux}
%-------------------------------------------------------------

The decomposition of vector fields (\ref{vec:dec}) requires an adaption of the path integral measure by a Jacobian $J_{\rm vec}$ such that
\be
\cD[C_\m] = J_{\rm vec}\; \cD[C^{\rm T}_\m]\cD[\eta] \, .
\ee
Following \cite{Mottola:1995sj}, the Jacobian can be found by explicit evaluation of a normalized Gaussian integral with the decomposition inserted,
\be \label{jac:vec}
\begin{split}
1 = &
\int \cD[C_\m]\;\; {\rm exp} \Big[ -\half \int d^dx\sqrt{g}\; C^\m C_\m \Big] \\
= &
J_{\rm vec}\; \int \cD[C^{\rm T}_\m]\cD[\eta]\;\; {\rm exp} \Big[ -\half \int d^dx\sqrt{g}\;\left(
C^{\rm T}{}^\m C^{\rm T}_\m + \eta \Dz \eta
\right) \\
\Rightarrow J_{\rm vec} = &
\bigg\{
\int\cD[\eta]\;\;{\rm exp}\Big[ -\half \int d^dx\sqrt{g}\;\eta \Dz \eta \Big]
\bigg\}^{-1} \, .
\end{split}
\ee
The remaining Gaussian integral gives a factor of $[\det{}_{(0)}(\Dz)]^{-\half}$ for bosonic or $\det{}_{(0)}(\D)$ for fermionic fields.

Likewise, the employment of the minimal TT-decomposition (\ref{mTTdec}) of the metric tensor requires to insert an appropriate Jacobian $J$. It contributes a compensating factor to the functional integral
\be
\cD[h_{\m\n}] = J_{\rm TT}\; \cD[h^{\rm T}_{\m\n}]\cD[h]\cD[\xi_\m] \, .
\ee
This Jacobian is also found in the same way by solving Gaussian integrals. We then find
\be
\begin{split}
1= & \; \int \cD[h_{\m\n}]\;\;
{\rm exp} \Big[-\half \int d^dx\sqrt{g}\; h^{\m\n} h_{\m\n} \Big] \\
= & \; J_{\rm TT} \;\int \cD[h^{\rm T}_{\m\n}]\cD[h]\cD[\xi_\m]\;\;
{\rm exp} \Big[-\half \int d^dx\sqrt{g}\;\left(
h^{\rm T}{}^{\m\n} h^{\rm T}_{\m\n} +\tfrac{1}{d} h^2 +2\xi_\m M^\m{}_\n \xi^\n
\right)\Big] \\
\propto & \; J_{\rm TT} \; \det{}_{(1)}(M)^{-\half} \, ,
\end{split}
\ee
with
\be \label{Mop}
M^\m{}_\n = \Dz \d^\m{}_\n -(1-\tfrac{2}{d}) D^\m D_\n -R^\m{}_\n \, .
\ee
Solving for $J_{\rm TT}$ leads to
\be\label{jac:mTT}
\begin{split}
J_{\rm TT} \propto & \det{}_{(1)}(M)^{\half} \,
= \, \det{}_{(1)}(M)^{-\half}\det{}_{(1)}(M)^{1} \\
= & \int \cD[\Y_\m]\cD[\bb_\m]\cD[b_\m]\;\;
{\rm exp} \Big[-\half \int d^dx\sqrt{g}\;\left(
\Y_\m M^\m{}_\n \Y^\n + \bb_\m M^\m{}_\n b^\n
\right)\Big] \, ,
\end{split}
\ee
where use has been made of the Faddeev-Popov trick to exponentiate $M^\m{}_\n$ introducing new auxiliary bosonic ($\Y$) and fermionic ($\bb$, $b$) vector fields.

%-------------------------------------------------------------
\section{Heat-kernel coefficients for fields with differential constraints}
\label{sect:heat}
%-------------------------------------------------------------
We now use the projection operators of App.\ \ref{sect:ttdec} to compute the early-time heat-kernel coefficients for transverse vectors and transverse-traceless matrices
on a generic compact manifold without boundaries. While the projectors are valid for any dimension $d$, we will restrict ourselves to the case $d=4$ for simplicity. The generalization to arbitrary $d$ is straightforward.

%-------------------------------------------------------------
\subsection{Standard Heat-kernel coefficients for unconstrained fields}
\label{App:B1}
%-------------------------------------------------------------
We start this section by summarizing the standard heat-kernel expansion of the minimal second order differential operator $\Delta = -D^2 - {\bf Q}$ for unconstrained fields. For small values $s$, the early-time expansion becomes
\be
\Tr_s \left[ \e^{-s\Delta} \right] = \frac{1}{(4\pi s)^2} \int d^4x \sqrt{g} \left[ \tr_s \, a_0 + s \, \tr_s \, a_2 + s^2 \, \tr_s \, a_4 + \cdots \right] \, ,
\ee
with coefficients \cite{Gilkey:1975iq}
\be
\begin{split}
a_0 = & \unit \, , \quad
a_2 = P \, , \quad \\
a_4 = & \frac{1}{180} \left( R_{\m\n\a\b} R^{\m\n\a\b} - R_{\m\n} R^{\m\n} +D^2 R \right) \unit + \frac{1}{2}P^2 + \frac{1}{12} \cR_{\m\n} \cR^{\m\n} + \frac{1}{6} D^2 P \, . 
\end{split}
\ee
Here the subscript $s=0,1,2$ indicates that the operator trace $\Tr_s$ acts on the space of scalars, vectors and symmetric matrices, respectively, while $\tr_s$ is a trace over the internal indices. Lastly, $P = \tfrac{1}{6} R \unit_s + {\bf Q}$ and $\cR_{\m\n} = 2 D_{[\m} D_{\n]}$.
For the special case where $\Delta = (-D^2 - q R) \unit_s$ the traced heat-kernel coefficients are
\be\label{heatcoeff}
\begin{array}{lll}
\tr_0 a_0 = 1  ,  & \tr_0 a_2 = \tfrac{1+6q}{6}  R \, , & \tr_0 a_4 = \tfrac{(1+6q)^2}{72}  R^2 + \tfrac{1}{180} (R_{\m\n\a\b} R^{\m\n\a\b} - R_{\m\n} R^{\m\n} ) \, , \\[1.1ex]
\tr_1 a_0 = 4  ,  & \tr_1 a_2 = \tfrac{2(1+6q)}{3}  R \, , & \tr_1 a_4 = \tfrac{(1+6q)^2}{18}  R^2 - \tfrac{1}{45} R_{\m\n} R^{\m\n} - \tfrac{11}{180} R_{\m\n\a\b} R^{\m\n\a\b} \, , \\[1.1ex]
\tr_2 a_0 = 10  ,  & \tr_2 a_2 = \tfrac{5(1+6q)}{3}  R \, , & \tr_2 a_4 = \tfrac{5(1+6q)^2}{36}  R^2 - \tfrac{1}{18} R_{\m\n} R^{\m\n} - \tfrac{4}{9} R_{\m\n\a\b} R^{\m\n\a\b} \, , \\ 
\end{array}
\ee
For $q=0$, these are summarized in the first three columns of Table \ref{t.heatcoeff}. These coefficients provide the starting ground for computing the heat-kernel coefficients for the constrained fields in the following.
%
%-------------------------------------------------------------
\subsection{Heat-kernel coefficients for transverse vector fields}
\label{App:B2}
%-------------------------------------------------------------
Our strategy for computing the heat-kernel for the constrained fields builds on the observation that it can be obtained from their unconstrained counterparts by inserting the suitable projection operators.
For the case of transverse vectors, the corresponding projector is given by \eqref{1Tproj}, yielding
\be\label{1Tmaster}
\begin{split}
\Tr_{\rm 1T}\left[ \e^{-s\Dz} \right] \equiv & \, \Tr_1 \left[ \e^{-s\Dz} \Pi_{\rm T} \right] \\ =  & \,  \Tr_1 \left[ \e^{-s\Dz} \right] + \Tr_0^\prime \left[ D_\mu \Dz^{-1} D^\mu \e^{-s\Dz} \right] \\
= & \, S_{1} + S_0 \, .  \\
\end{split}
\ee
Here, the prime on the scalar trace highlights that the constant scalar mode does not contribute, since it drops out from the decomposition \eqref{vec:dec}.

The expansion of the trace $S_1$ can be directly read off from \eqref{heatcoeff}. For the evaluation of $S_0$, we first calculate the full trace, before dealing with the constant mode. The expansion of the trace can be found by first commuting $D^\m$ with $\e^{-s\D}$. Using \eqref{comexp} together with \eqref{exactcom} and \eqref{doublecom} acting on scalars and exploiting the cyclicity of the trace, this results in
\be
\begin{split}
S_0 = & \, \Tr_0 \left[ D_\mu \Dz^{-1} \e^{-s\Dz} \left( D^\mu + s R^{\m\n} D_\n + \half s^2 R^{\m\a} R_{\a\n} D^\n \right) \right]
+ \cO(R^3) \, \\
= & \, - \Tr_0\left[ \e^{-s\Dz} \right] + s \Tr_0 \left[ \Dz^{-1} \e^{-s\Dz} R^{\m\n} D_\m D_\n \right] + \tfrac{s^2}{2} \Tr_0 \left[ \Dz^{-1} \e^{-s\Dz} R^{\m\a} R_{\a\n} D_\mu D^\n \right] \\
= & \, T_0 + T_1 + T_2 + \cO(R^3)\, .
\end{split}
\ee

$T_0$ is given by the standard heat-kernel expansion of a scalar field. To evaluate $T_1$ and $T_2$ we combine the Schwinger-trick with the off-diagonal heat-kernel methods of Section \ref{sect:2.3},
\be\label{Ansexample}
\begin{split}
T_1 = & \, s \, \Tr_0 \left[ \Dz^{-1} \e^{-s\Dz} R^{\m\n} D_\m D_\n \, \right] \, \\
= & \, s \, \int_0^\infty dt \, \Tr_0 \left[ \e^{-(s+t) \D} \, R^{\m\n} D_\m D_\n \, \right] \\
= & \, s \, \int_0^\infty dt \, \tfrac{1}{(4\pi(s+t))^2} \, \int d^4x \sqrt{g} \, R^{\m\n} 
\, \left[ - \tfrac{1}{2(s+t)} \, g_{\m\n} \, \left(1 + \tfrac{1}{6} (s+t) R \right) + \tfrac{1}{6} R_{\m\n} \right] \\
= & \, \tfrac{1}{(4\pi s)^2} \, \int d^4x \sqrt{g} \, \left[
- \frac{1}{4} s R - \frac{1}{12} s^2 R^2 + \frac{1}{6} s^2 R_{\m\n} R^{\m\n} 
\right] \, ,
\end{split}
\ee
where we substituted the $H_{\m\n}$-tensor \eqref{atexplicit} in the third step.
Following the same route, one finds
\be
\begin{split}
T_2 = & \, \tfrac{1}{(4\pi s)^2} \int d^4x \sqrt{g} \left[ - \tfrac{1}{8} s^2 R_{\m\n} R^{\m\n} 
\right] \, .
\end{split}
\ee

In order to obtain the final result, we still have to deal with the zero mode appearing in $S_1$. Its subtraction from $T_0$ is trivial, using the identity $\Tr^\prime_0 = \Tr_0 - T^{\rm zero-mode}_0$, where $T^{\rm zero-mode}_0 = -1$ is a pure number, counting the constant scalar field on the manifold. In particular, it is independent of the volume-element. Thus, it has to contribute to an index, which in the present case is given by the Euler-character $\chi_{\rm E}$ of the manifold $\int d^4x \sqrt{g} (R^2 - 4 R_{\m\n} R^{\m\n} + R_{\m\n\a\b} R^{\m\n\a\b}) = 32 \pi^2 \chi_{\rm E}$. The zero-mode then gives a contribution to the heat-kernel expansion proportional to the inverse Euler-character,
\be\label{zmcont}
T^{\rm zero-mode}_0 = - \frac{1}{(4\pi)^2} \frac{1}{2 \chi_{\rm E}} \int d^4x \sqrt{g} (R^2 - 4 R_{\m\n} R^{\m\n} + R_{\m\n\a\b} R^{\m\n\a\b}) \, .
\ee
For the other terms we have to deal with undefined expressions (zero over zero), so we need to regularize them.
We take such a regularization to be a sequence (continuous or discrete) of scalar modes $\phi^{(m)}$ such that $lim_{m\to 0}\phi^{(m)} = \phi^{(0)} = \text{const.}$, and such that for any $m>0$ there exists a well defined tensor $s^{(m)}_{\m\n}$ for which
\be
\Dz^{-1} D_\m D_\n \phi^{(m)} = s^{(m)}_{\m\n}  \phi^{(m)} \, .
\ee
By counting the number of derivatives we assume that for a suitable regularization the limit $s^{(0)}_{\m\n} = \lim_{m\to 0} s^{(m)}_{\m\n} $ will exist, and it should satisfy $D_\r s^{(0)}_{\m\n} = 0$, together with $g^{\m\n} s^{(0)}_{\m\n} = -1$.
There is only one tensor satisfying these requirements, namely
\be
s^{(0)}_{\m\n} = - \tfrac{1}{d} g_{\m\n} \, .
\ee
Comparing these contributions to \eqref{zmcont} we observe that they are of higher order in the curvature, and thus cannot contribute to the heat-coefficients computed here.

Combining all these partial results via \eqref{1Tmaster} yields the heat-kernel coefficients for the transverse vector-fields
\be \label{1Theat}
\begin{split}
\tr_{\rm 1T} a_0 = & \, 3 \, , \quad \tr_{\rm 1T} a_2 = \tfrac{1}{4} R \, , \quad \\ 
\tr_{\rm 1T} a_4 = & \, \left( - \tfrac{1}{24} + \tfrac{1}{2\chi_{\rm E}} \right) R^2 + \left( \tfrac{1}{40} - \tfrac{2}{\chi_{\rm E}} \right) R_{\m\n} R^{\m\n} - \left( \tfrac{1}{15} - \tfrac{1}{2\chi_{\rm E}} \right) R_{\m\n\rho\sigma} R^{\m\n\rho\sigma} \, . 
\end{split}
\ee
This proves the fourth column of Table \ref{t.heatcoeff}.

%-------------------------------------------------------------
\subsection{Heat-kernel coefficients for transverse-traceless matrices}
\label{App:B3}
%-------------------------------------------------------------
The computation of the heat-kernel coefficients in the 2T-case is completely analogous to the 1T-case, but technically more involved. Using the projectors \eqref{2Tprojector} and \eqref{2Tproj} we find
\be\label{2Tmaster}
\begin{split}
\Tr_{\rm 2T} \left[ \e^{-s\Dz} \right] \equiv 
\Tr_{\rm 2} \left[ \e^{-s\Dz} \Pi_{\rm 2T} \right] =  & \, 
\Tr_{\rm 2} \left[ \e^{-s\Dz} \right] - \Tr_{\rm 2} \left[ \e^{-s\Dz} \Pi_{\rm 2L} \right] - \Tr_{\rm 2} \left[ \e^{-s\Dz} \Pi_{\rm tr} \right] \, \\
=& \, S_2 + S_1 + S_0 \, .
\end{split} 
\ee
Noting that
\be
S_0 = -\Tr_{\rm 2} \left[ \, \e^{-s\Dz} \, \Pi_{\rm tr} \, \right] = -\Tr_{\rm 0} \left[ \e^{-s\Dz} \right] \, ,
\ee
both $S_2$ and $S_0$ reduce to standard heat-kernel formulas and their contribution can be read off from \eqref{heatcoeff}.

In order to obtain $S_1$, we first substitute $\Pi_{\rm 2L}$, \eqref{2Tproj} and use the cyclicity of the trace to write
\be\label{longTr}
\begin{split}
S_1 &= \Tr_{\rm 1} \left[  (-D^\g) \left[ \unit_2 - \Pi_{\rm tr} \right]_{\g\b}{}^{\m\n} \e^{-s\D}  \left[ P_1 \right] _{\m\n}^\a \left[ P_2^{-1} \right]_\a^{\b'} \right] \\
 & = \Tr_{\rm 1} \left[   \e^{-s\D}\right] 
+\Tr_{\rm 1} \left[  -s \e^{-s\D} [D^\g,D^2] \left[P_1 \right]_{\g\b}^\a \left[ P_2^{-1}\right]_\a^{\b'} \right]\\
 & \quad +\Tr_{\rm 1} \left[  -\frac{s^2}{2} \e^{-s\D} \left[[D^\g,D^2],D^2\right] \left[ P_1 \right] _{\g\b}^\a \left[ P_2^{-1}\right]_\a^{\b'} \right] + \cO(R^3)\\
 & \equiv T_0 + T_1 + T_2 + \cO(R^3) \, .
\end{split}
\ee
The commutator in the second step follows from \eqref{comexp} and we exploited the orthogonality of the projectors, which ensures $\Pi_{\rm tr} \cdot P_1 = 0$. The commutators \eqref{exactcom}, \eqref{doublecom} and the perturbative inverse $P_2^{-1}$, \eqref{cres}, can then be explicitly substituted. Expanding the products, the full trace is then built from linear combinations of basis integrals, which can be evaluated with the off-diagonal heat-kernel along the lines of \eqref{Ansexample}. For convenience, these are summarized in Table \ref{t.Abasisint}.

\begin{table}[t!]
\begin{center}
\begin{tabular}{lcl}\hline
$s \Tr_0 \left[ \frac{1}{\Dz} \e^{-s \Dz} R^2 \right] $& $=$ & $\frac{1}{(4\pi)^2} \int d^4x \sqrt{g} R^2 $ \\
$s \Tr_0 \left[ \frac{1}{\Dz} \e^{-s \Dz} R^{\m\n} D_\m D_\n \right]$ & $ = $
& $\frac{1}{(4\pi)^2} \int d^4x \sqrt{g} \left[ - \frac{1}{4s}R - \frac{1}{12}R^2 + \frac{1}{6} R_{\m\n}R^{\m\n} \right] $  \\
$s \Tr_0 \left[ \frac{1}{\Dz^2} \e^{-s \Dz} R^{\m\a} R_\a{}^\n D_\m D_\n \right] $ & $
 = $ &  $  \frac{1}{(4\pi)^2} \int d^4x \sqrt{g} \left[ -\frac{1}{4} R^{\m\n} R_{\m\n} \right]$ \\
 $s \Tr_0 \left[ \frac{1}{\Dz^2} \e^{-s \Dz} R R^{\m\n} D_\m D_\n \right]
 $ & $= $ & $\frac{1}{(4\pi)^2} \int d^4x \sqrt{g} \left[ -\frac{1}{4} R^2 \right] $  \\
$s \Tr_0 \left[ \frac{1}{\Dz^3} \e^{-s \Dz} R^{\m\n} R^{\a\b} D_\m D_\n D_\a D_\b \right]
 $ & $= $ & $  \frac{1}{(4\pi)^2} \int d^4x \sqrt{g} \left[ \frac{1}{24} R^2 + \frac{1}{12} R^{\m\n} R_{\m\n} \right] $ \\
$ s \Tr_0 \left[ \frac{1}{\Dz^2} \e^{-s \Dz}  R_{\a\b} R^{\a\m\n\b} D_\m D_\n \right] $
 & $ = $ & $  \frac{1}{(4\pi)^2} \int d^4x \sqrt{g} \left[ \frac{1}{4} R^{\m\n} R_{\m\n} \right] $ \\ \hline 
\end{tabular}
\caption{\label{t.Abasisint} Typical operator traces arising in the computation of perturbed traces. These are evaluated using the off-diagonal heat-kernel methods discussed in Subsection \ref{sect:2.3}.}
\end{center}
\end{table}

At this point, the following technical remark is in order. When evaluating operator traces with the off-diagonal heat-kernel it is {\it crucial} that all covariant derivatives appear in a totally symmetric way. This requires their symmetrization via $D_\m D_\n = D_{(\m} D_{\n)} + \half [D_\m, D_\n] $, with the antisymmetric part expressed as curvature tensors. In this respect it is important to note that the open vector indices $\beta$ and $\beta'$ appearing in \eqref{longTr} may not be contracted immediately. Instead, $\beta'$ must be thought of as contracted with a vector $\phi_{\b'}$ to the right of the expression. Only once all commutators have been expressed through curvature tensors, the ``open indices'' $\beta$ and $\beta'$ may be contracted.

Following this route, a lengthy but straightforward computation yields
\be
T_1 = \frac{1}{(4 \pi s)^2} \int d^4x \sqrt{g} \left[ \frac{5}{3} s R + \frac{19}{27} s^2 R^2 - \frac{22}{27} s^2 R_{\m\n} R^{\m\n} - \frac{4}{3} s^2 R_{\m\n\rho\sigma}
R^{\m\n\rho\sigma}
\right] \, ,
\ee
and
\be
T_2 = \frac{1}{(4 \pi s)^2} \int d^4x \sqrt{g} \left[ \frac{17}{18} s^2 R_{\m\n} R^{\m\n} + \frac{2}{3} s^2 R_{\m\n\rho\sigma} R^{\m\n\rho\sigma}  \right] \, .
\ee
Combining this result with the standard vector-trace $T_0$, we arrive at the vector contribution to the 2T-trace,
\be\label{vec:trace}
S_1 = \frac{1}{(4\pi s)^2} \int d^4x \sqrt{g}
\left[
4 + \tfrac{7}{3} s R + \tfrac{41}{54} s^2 R^2 + \tfrac{29}{270} s^2 R_{\m\n} R^{\m\n} - \tfrac{131}{180} s^2 R_{\m\n\rho\sigma} R^{\m\n\rho\sigma} 
\right]\, .
\ee

As observed above, $S_2$ and $S_0$ are standard heat-traces involving symmetric matrices and scalars, respectively, so that they can directly be inferred from \eqref{heatcoeff}. Adding \eqref{vec:trace}, the heat-kernel coefficients
for the 2T-trace are obtained as
\be\label{2Theat}
\begin{split}
\tr_{\rm 2T} a_0 =  5 \, , \quad
\tr_{\rm 2T} a_2 =  - \tfrac{5}{6} \, R \, , \quad
\tr_{\rm 2T} a_4 =  - \tfrac{137}{216} R^2 - \tfrac{17}{108} R_{\m\n} R^{\m\n} + \tfrac{5}{18} R_{\m\n\rho\sigma} R^{\m\n\rho\sigma} \, .
\end{split}
\ee
This constitutes the main result of this subsection and proves the fifth column of Table \ref{t.heatcoeff}.

We close this subsection with a remark on the zero modes, which do not contribute to the minimal TT-decomposition and have to be subtracted from the coefficients.
Inspecting \eqref{mTTdec}, we conclude that vector fields satisfying the conformal Killing equation
\be\label{2T:zeromodes}
D_\m \xi_\n + D_\n \xi_\m - \tfrac{1}{2} g_{\m\n} D^{\a} \xi_\a = 0 \,
\ee
will not contribute to the fluctuation field, and therefore have to be excluded from the operator traces. The number of modes to be excluded from the trace is then given by the number of Killing vectors $n_{\rm KV}$ (satisfying $D_\m \xi_\n + D_\n \xi_\m = 0$)
plus the number of conformal Killing vectors $n_{\rm CKV}$ which solve \eqref{2T:zeromodes} with $D^{\a} \xi_\a^{\rm CKV} \not = 0$. Following the logic for the zero modes of the previous subsection, these modes give rise to a correction of $S_1$ proportional to the Euler integrand
\be\label{symextra}
\tr_{\rm 2T} a_4^{\rm symmetry} = \frac{N}{2\chi_{\rm E}} \left( R^2 - 4 R_{\m\n}R^{\m\n} + R_{\m\n\a\b} R^{\m\n\a\b} \right) \, , 
\ee
where $N = n_{\rm KV} + n_{\rm CKV} $ is the number of independent solutions of \eqref{2T:zeromodes}.
This extra contribution is absent in the generic case where the manifold does not posses particular symmetries. However, it plays an important role when comparing our general computation to the spherical result in the next subsection.

%-------------------------------------------------------------
\subsection{Comparison of the Heat-kernel coefficients on a spherical background}
\label{App:B4}
%-------------------------------------------------------------
An important check for the heat-kernel coefficients \eqref{1Theat} and \eqref{2Theat} is provided by the special case where $g_{\m\n}$ is the metric of the four-sphere $S^4$. In this case, the coefficients shown in Table \ref{t.heatcoeff} have to reproduce the earlier results \cite{Lauscher:2001ya,Codello:2008vh}. In this subsection, we verify that this is indeed the case.

Since $S^4$ is a maximally symmetric space, all its curvature tensors can be expressed in terms of the metric and the covariantly constant Ricci-scalar
\be\label{geo:sphere}
R_{\m\n} = \frac{1}{4} \, R \, g_{\m\n}, \qquad R_{\m\n\r\s} = \frac{1}{12} \, R \, ( g_{\m\r} g_{\n\s}- g_{\m\s} g_{\n\r}).
\ee
These relations imply that the curvature-square and volume-terms can be expressed in terms of the Ricci-scalar,
\be\label{geo:id}
R_{\m\n} R^{\m\n} = \frac{1}{4} R^2 \, , \quad R_{\m\n\r\s} R^{\m\n\r\s} = \frac{1}{6}\, R^2 \, , \quad
{\rm Vol}_{S^4} = 384 \, \pi^2 \, R^{-2} \, .
\ee
Moreover, the four-sphere has $\chi_{\rm E} = 2$ and admits $n_{\rm KV} = 10$ Killing- and $n_{\rm CKV} = 5$ conformal Killing vectors, totallying to $N=15$. Substituting this data into \eqref{1Theat} and \eqref{2Theat}, taking into account the contributions arising on a special background with symmetries \eqref{symextra}, one obtains
\be\label{heatsphere}
\begin{array}{lll}
\left. \tr_{\rm 1T} a_0 \right|_{S^4} = 3 \, , \quad & 
\left. \tr_{\rm 1T} a_2 \right|_{S^4} = \tfrac{1}{4} R \, , \quad &
\left. \tr_{\rm 1T} a_4 \right|_{S^4} = -\tfrac{7}{1440}R^2\, , \\[1.2ex]
\left. \tr_{\rm 2T} a_0 \right|_{S^4} =  5 \, , \quad &
\left. \tr_{\rm 2T} a_2 \right|_{S^4} =  - \tfrac{5}{6} \, R \, , \quad & 
\left. \tr_{\rm 2T} a_4 \right|_{S^4} = -\tfrac{1}{432}R^2 \,.
\end{array}
\ee
These are exactly the tabulated results given in \cite{Lauscher:2001ya,Codello:2008vh}. We thus establish that the projector-method employed here recovers the heat-kernel results obtained previously.

%-------------------------------------------------------------
\section{Decoupling gauge degrees of freedom, ghosts, and auxiliaries }
\label{sect:dec}
%-------------------------------------------------------------
The use of standard gauge fixing terms and field decompositions leads to recurring terms in the RG equation that can be evaluated independently of the specific model under consideration.
Here, we will demonstrate that it is possible to separate the contributions of the gauge fixing and Faddeev-Popov terms to the running of coupling constants from those of the physical part of the action.
To achieve this, a gauge field $A$ is first decomposed in background and quantum fluctuations as $A=\bar{A} +a$, and the quantum fields are
then split into component fields such that the gauge-fixing condition contains only certain modes, denoted as $a_{\rm g}$.
With an appropriately chosen gauge fixing condition $F(\bar{A},a)=F(\bar{A},a_{\rm g})$ linear in the fluctuation field, the gauge fixing term can be written in the form
\be \label{Sgf}
\begin{split}
\G^{\rm gf}_k \,
=& \, \frac{1}{2\a}\int d^dx\sqrt{\gb}\; F(\bar{A},a)^2 \\
=& \, \frac{1}{2\a}\int d^dx\sqrt{\gb}\; a_{\rm g} G(\bar{A}) a_{\rm g} \, ,
\end{split}
\ee
for some operator $G$ depending on the background fields only. It should be noted that invariance with respect to the background gauge transformation can always be maintained irrespective of the separation of modes.

In order to write the structure of the RG equation in terms of the decomposed gauge fields, we can arrange the latter in a multiplet $\Phi=(\phi,a_{\rm g})$, where $\phi$ represents all remaining component fields of $A$ and other fields occurring in the action like the Faddeev-Popov ghosts. Any of these fields may be later further decomposed into component fields.

With the gauge part formally separated, the quadratic part of the average effective action assumes the form
\be
\G^{\rm quad}_k =
\frac{1}{2}\Phi\left[
\begin{array}{cc}
L & Q \\
\Qt & \tfrac{1}{\a}G+H
\end{array}
\right]\Phi \, ,
\ee
where the block structure is chosen such that the lower right is the quadratic part of the pure gauge component, $L$ denotes the contributions from all remaining fields, and $Q$ and $\Qt$ are mixed terms. Schematically, the FRGE \eqref{FRGE} then becomes
\be\label{D.3}
\p_t \G_k =
\frac{1}{2}{\rm STr}
\left[
\begin{array}{cc}
L+R_L & Q+R_Q \\
\Qt+R_{\Qt} & \tfrac{1}{\a}(G+R_G)+H+R_H
\end{array}
\right]^{-1}
\p_t\left[
\begin{array}{cc}
R_L & R_Q \\
R_{\Qt} & \tfrac{1}{\a}R_G+R_H
\end{array}
\right] \, .
\ee
The crucial thing to note here is that a suitably constructed cutoff for the gauge component $R_G$ according to the scheme (\ref{typeIcutoff}) must come with the same $\alpha$-dependence as $G$. In \eqref{D.3}, this feature is highlighted by displaying the $\alpha$-dependence explicitly.
A physically motivated choice of the gauge parameter is that of the Landau-de Witt gauge ($\a=0$), since there is no smearing of the gauge condition. Moreover, $\a=0$ is expected to be a fixed point for the gauge parameter \cite{Ellwanger:1995qf,Litim:1998qi}. Thus $\a = 0$ is a natural choice. The occurrence of $\a^{-1}$ in the cutoff term does not allow to take the limit $\a\rightarrow 0$ right away, however. Rather, we must keep track of $\a$ in linear order to cancel the inverse.

With the help of the general expression for the inverse of a block matrix (\ref{Ginv}), we can expand the inversion around $\a=0$ up to linear order, yielding
\be\label{ginv:lin}
\left[
\begin{array}{cc}
L & Q \\
\Qt & \tfrac{1}{\a}G+H
\end{array}
\right]^{-1}
=\left[\begin{array}{cc} L^{-1} & 0 \\ 0 & 0 \end{array}\right]
+\a \left[\begin{array}{cc} L^{-1} Q G^{-1} \Qt L^{-1} & -L^{-1} Q G^{-1} \\ -G^{-1} \Qt L^{-1} & G^{-1} \end{array}\right]
+\cO(\a^2) \, .
\ee
This formula requires the invertability of $L$ and $G$, a fact which we have to keep in mind in the explicit construction of the gauge fixing term.

Multiplying \eqref{ginv:lin} with the cutoff matrix and taking the limit $\a\rightarrow 0$, we find
\be
\begin{split}
\p_t \G_k =&
\frac{1}{2}{\rm STr}
\left[
\begin{array}{cc}
(L+R_L)^{-1}\p_t R_L & 0 \\
0 & (G+R_G)^{-1}\p_t R_G
\end{array}
\right]
+\cO(\a) \\
=&
\frac{1}{2}{\rm STr} \left[ (L+R_L)^{-1}\p_t R_L \right] + \frac{1}{2}{\rm STr} \left[ (G+R_G)^{-1}\p_t R_G \right] \, .
\end{split}
\ee
In the Landau gauge limit, the structure of the RG equation simplifies such that the mixed terms $Q$ and $\Qt$ as well as the block $H$ drop out identically. Therefore, the pure gauge components decouple from the remaining fields and their contribution is determined by the gauge fixing term only. This feature will be exploited when computing the universal contribution to the gravitational RG flow in the next section.

Furthermore, a standard ghost term of the form
\be \label{Sgh}
\begin{split}
\G^{\rm gh}_k
=& \int d^dx \sqrt{\gb} \, \Cb \frac{\delta F}{\delta a} \delta_C \, A  
= \int d^dx \sqrt{\gb} \, \Cb \, M \, C
\end{split}
\ee
is needed in the case of non-Abelian gauge theories. Because of its simple quadratic form, it is possible to examine the structure of $L$ by separating the ghost fields from the remaining ones. Making use of the background independence of the RG equation, we can set the background ghost field to zero, eliminating the cross-terms stemming from the ghost-gauge field interaction. Note that this choice will of course not allow us to extract the running of the coupling constants in the ghost sector. The block structure then becomes 
\be
L=
\left[\begin{array}{ccc}
K & 0 & 0 \\
0 & 0 & -M \\
0 & M & 0
\end{array}\right] \, ,
\ee
where $M$ is the kernel of the ghost action (\ref{Sgh}) and $K$ is the still undetermined block containing all other second variations of the action. Since the cutoff $R_L$ must be of the same form, the resulting trace in the RG equation decomposes as
\be
\frac{1}{2}{\rm STr} (L+R_L)^{-1}\p_t R_L =
\frac{1}{2}{\rm Tr} (K+R_K)^{-1}\p_t R_K - {\rm Tr} (M+R_M)^{-1}\p_t R_M \, ,
\ee
and thus
\be
\p_t \G_k =
\frac{1}{2}{\rm Tr} (K+R_K)^{-1}\p_t R_K + \frac{1}{2}{\rm Tr} (G+R_G)^{-1}\p_t R_G - {\rm Tr} (M+R_M)^{-1}\p_t R_M \, .
\ee

In order to arrive at an entirely self-contained treatment of those degrees of freedom, it is necessary to include a sector of auxiliary fields that formally correct the counting of zero modes of the applied decompositions. Using the Faddeev-Popov trick, these terms are found by exponentiating the Jacobian created by the corresponding projection operators. Since these operators always act linearly, the resulting terms will be strictly quadratical in the auxiliary fields and decouple from the other fields like the ghost sector by choosing a zero background.

Therefore, the gauge and ghost as well as the auxiliary terms decouple from the physical field components in the Landau-de Witt gauge. This result allows us to evaluate their contribution to the RG flow completely independently, leaving the matter field content of the model arbitrary, and to give universal expressions to be reused in a broad class of models.

%-------------------------------------------------------------
\section{Universal contributions to the gravitational RG flow}
\label{App:E}
%-------------------------------------------------------------
In this appendix, we compute the re-occurring contributions $\cS^{\rm uni}$, arising from the gauge fixing, ghost, and auxiliary actions, up to linear order in the background curvature. These naturally organize themselves according to
\be\label{orguniv}
\cS^{\rm uni} = \cS^{\rm gf} + \cS^{\rm gh} + \cS^{\rm aux} \,,
\ee
and we will evaluate these traces in turn. Again, we set $d=4$ for simplicity, but the generalization to arbitrary $d$ is straightforward.

%-------------------------------------------------------------
\subsection{Gauge fixing the gravitational action}
\label{App:E1}
%-------------------------------------------------------------
The construction of the effective average action employs the background field method, which breaks the quantum gauge transformations, but retains background gauge-transformations as an auxiliary symmetry (see \cite{Niedermaier:2006wt} for more details).
Neglecting the $k$-dependence, this can be achieved by adding the classical gauge-fixing term
\be\label{S.gf}
S^{\rm gf} = \frac{1}{2\alpha} \int d^4x \sqrt{\gb} \, \gb^{\m\n} F_\m F_\n \, , \qquad F_\m =  \Db^\n h_{\m\n} - \beta \Db_\m h
\ee
to the gravitational action. Notably, $S^{\rm gf}$ is quadratic in the fluctuation field $h_{\m\n}$. Substituting the minimal-TT-decomposition \eqref{mTTdec}, we find
\be \label{gaugechoice}
F_\m = \Db^2 \xi_\mu + \tfrac{1}{2} \Db_\mu \Db^\nu \xi_\nu + \Rb_{\m}{}^{\n} \xi_\n +(\tfrac{1}{4}-\beta) \Db_\m h \, ,
\ee
where $\beta$ is a freely adjustable gauge parameter. To ensure the invertibility of the Hessian from the gauge fixing sector, $\beta = 1/4$ has to be chosen. In this case, all gauge degrees of freedom are captured by the vector $\xi_\m$. By virtue of the decoupling theorem of App.\ \ref{sect:dec}, this field decouples from the gravitational part of the action once the Landau limit $\alpha \rightarrow 0$ is invoked.

Setting $\beta = 1/4$ and dropping the bar from background quantities to simplify the notation, eq.\ \eqref{S.gf} becomes
\be \label{gffactor}
\begin{split}
S^{\rm gf}
=& \frac{1}{2\alpha} \int d^4x \sqrt{g} \, \xi_\mu
\left[ \delta^\m{}_\a \Dz - \tfrac{1}{2} D^\m D_\a - R^\m{}_\a \right] \,
\left[ \delta^\a{}_\n \Dz - \tfrac{1}{2} D^\a D_\n - R^\a{}_\n \right] \xi^\nu \, .
\end{split}
\ee
Following Sect.\ \ref{sect:algo}, we eliminate the differential operators of non-minimal form by the transverse decomposition of the vector $\xi_\m = \xi^{\rm T}_\m + D_\m \sigma$ with $D^\m \xi^{\rm T}_\m = 0$. In terms of these component fields the action becomes
\be\label{SgfTT}
\begin{split}
S^{\rm gf}
= \frac{1}{2\alpha} \int d^4x \sqrt{g}\; \Big\{ &
  \xi^{\rm T}_\m \big[ \Kbb_{\rm 1T}^{\rm gf} + \Mbb_{\rm 1T}^{\rm gf} \big]^\m{}_\n \xi^{\rm T}{}^\n
+ \sigma \big[ \Kbb_{\rm 0}^{\rm gf} + \Mbb_{\rm 0}^{\rm gf} \big] \sigma \\ & \quad
+ \xi^{\rm T}_\m \big[ \Mbb_{\times}^{\rm gf} \big]^\m  \sigma
+ \sigma \big[ \Mbb_{\times}^{\rm gf} \big]_\n^\dagger \xi^{\rm T}{}^\n
\Big\} \, ,
\end{split}
\ee
with kinetic terms
\be \label{gf:kin}
\begin{split}
\big[ \Kbb_{\rm 1T}^{\rm gf} \big]^\m{}_\n = \Dz^2 \delta^\m{}_\n \, , \qquad
\big[ \Kbb_{\rm 0}^{\rm gf} \big] = \tfrac{9}{4} \D^3 \, ,
\end{split}
\ee
and vertices
\be \label{gfvertex}
\begin{array}{lcl}
\big[ \Mbb_{\rm 1T}^{\rm gf} \big]^\m{}_\n & = & - R^\m{}_\n \Dz - \Dz R^\m{}_\n + R^{\m\a} R_{\a\n} \, , \\
\big[ \Mbb_{\rm 0}^{\rm gf} \big] & = & 3 \left( \D D_\m R^{\m\n} D_\n + D_\m R^{\m\n} D_\n \D \right) - 4 D_\m R^{\m\a} R_{\a}{}^\n D_\n \, , \\
\big[ \Mbb_{\times}^{\rm gf} \big]^\m & = & \, - 3 R^{\m\n} D_\n \D - 2 \D R^{\m\n} D_\n + 2 R^{\m\n} R_{\n}{}^\a D_\a  \, , \\
\big[ \Mbb_{\times}^{\rm gf} \big]_\n^\dagger & = & \, 3 \Dz D^\m R_{\m\n} + 2 D^\m R_{\m\n} \Dz - 2 D^\m R_{\m\a} R^\a{}_{\n} \, .
\end{array}
\ee
Hence, the gauge fixing sector of $\Gamma^{(2)}_k$ assumes a two-by-two block structure whose entries do not contain $\Dbb$-terms.

Following the rule \eqref{TypeIb} of regularizing the kinetic terms, the cutoff-operators become
\be
\cR_k^{\rm gf,1T} = \delta^\m{}_\n (P_k^2-\Dz^2) \, , \qquad
\cR_k^{\rm gf,0} = \tfrac{9}{4} (P_k^3-\Dz^3) \, .
\ee
The corresponding regularized kinetic terms are denoted by $\Pbb_i^{\rm gf} = \Kbb_i^{\rm gf} + \cR_k^{{\rm gf}, i}$, where $i = 0, {\rm 1T}$.

The final step consists of evaluating the resulting operator traces along the lines of Sect.\ \ref{sect:algo}. Since $\Mbb_{\times}$ and $\Mbb_{\times}^\dagger$ are both linear in $R$, the cross-terms do not contribute to the present truncation. Terminating the perturbative inversion at linear order in $\Mbb$, the ${\rm 1T}$-sector evaluates to
\be \label{gfvector}
\begin{split}
\cS_{\rm 1T}^{\rm gf} = & \, \half \Tr_1 \left[ \frac{1}{\Pbb_{\rm 1T}^{\rm gf}} \p_t \cR_k^{\rm gf,1T} \Pi_{\rm T} \right]
- \half \Tr_1 \left[ \frac{1}{\Pbb_{\rm 1T}^{\rm gf}} \Mbb_{\rm 1T}^{\rm gf} \cdot \Pi_{\rm T}  \frac{1}{\Pbb_{\rm 1T}^{\rm gf}} \p_t \cR_k^{\rm gf,1T} \right] \\
= & \, \frac{1}{(4\pi)^2} \int d^4x \sqrt{g} \, \left[ 3 \, Q_2[f^1] + R \left( \tfrac{1}{4} Q_1[f^1] + 3 Q_3[f^3] \right) \right] \, ,
\end{split}
\ee
while the scalar sector gives
\be
\begin{split}
\cS_{\rm 0}^{\rm gf} = & \, \half \Tr_0 \left[ \frac{1}{\Pbb_{\rm 0}^{\rm gf}} \p_t \cR_k^{\rm gf,0} \right] - \half \Tr_0 \left[ \frac{1}{\Pbb_{\rm 0}^{\rm gf}} \Mbb_{\rm 0}^{\rm gf} \frac{1}{\Pbb_{\rm 0}^{\rm gf}} \p_t \cR_k^{\rm gf,1T} \right] \\
= & \, \frac{1}{(4\pi)^2} \int d^4x \sqrt{g} \, \left[ \tfrac{3}{2} \, Q_2[f^1] + R \left( \tfrac{1}{4} Q_1[f^1] + 6 Q_4[f^4]\right) \right] \, .
\end{split}
\ee
Here the argument of the $Q$-functionals is defined as
\be \label{fdef}
f^n = \frac{1}{(P_k)^n} \p_t R_k \, .
\ee
The complete contribution of the gauge-sector is then given by
\be \label{Sgftotal}
\begin{split}
\cS^{\rm gf} = & \cS_{\rm 1T}^{\rm gf} + \cS_{\rm 0}^{\rm gf} \\
=& \frac{1}{(4\pi)^2} \int d^4x \sqrt{g} \, \left[ \tfrac{9}{2} \, Q_2[f^1] + R \left( \tfrac{1}{2} Q_1[f^1] + 3 Q_3[f^3] + 6 Q_4[f^4]\right) \right] \, .
\end{split}
\ee

%-------------------------------------------------------------
\subsection{The ghost sector}
%-------------------------------------------------------------
The ghost term corresponding to the gauge fixing (\ref{gaugechoice}) with $\beta = 1/4$ is given by
\be\label{Sghost}
\begin{split}
S_{\rm gh} & =  \, \int d^4x \sqrt{\gb} \, \Cb_\m \, \gb^{\m\n} \, \frac{\delta F_\n}{\delta h_{\a\b}} \, \cL_C(\gb + h)_{\a\b}
= \, \int d^4x \sqrt{\gb} \, \Cb_\m \, \cM^{\m}{}_{\n} \, C^{\n} \, ,
\end{split}
\ee
with the Faddeev-Popov determinant
\be
\cM^{\m}{}_{\n} = \gb^{\m\a} \gb^{\lambda\sigma} \left[ \Db_\lambda g_{\a\n} D_\sigma + \Db_\lambda g_{\sigma\n} D_\a  - \half \Db_\a g_{\lambda\n} D_\sigma \right] \, ,
\ee
Its part quadratic in the fluctuation fields is
\be\label{ghostquad}
S_{\rm gh}^{\rm quad} =  \, \int d^4x \sqrt{g} \, \Cb_\m \, \left[ \delta^\m{}_\n \Dz - \half D^\m D_\n - R^\m{}_\n \right] \, C^\n \, ,
\ee
where we again identified $g = \gb$. Since the operator \eqref{ghostquad} still contains $\Dbb$-type terms, we carry out the transverse decomposition of the anti-ghost field $\Cb_\m = \Cb_\m^{\rm T} + D_\m \bar{\eta}$ and likewise for the ghost. The resulting $S_{\rm gh}^{\rm quad}$ assumes a two-by-two block form without $\Dbb$-type operators,
\be\label{ghdec}
S_{\rm gh}^{\rm quad} =  \, \int d^4x \sqrt{\gb} \, \Big\{
 \Cb_\m^{\rm T} \left[ \Kbb_{\rm 1T}^{\rm gh} + \Mbb_{\rm 1T}^{\rm gh} \right]^\m_\n C^{{\rm T} \n}  
 + \bar{\eta} \left[ \Kbb_{\rm 0}^{\rm gh} + \Mbb_{\rm 0}^{\rm gh} \right] \eta
 + \bar{\eta} \left[ \Mbb_{\times}^{\rm gh} \right]_\n^\dagger C^{{\rm T} \n} + \Cb_\m^{\rm T} \left[ \Mbb_{\times}^{\rm gh} \right]^\m \eta \Big\} ,
\ee
with kinetic terms
\be\label{ghkin}
\big[ \Kbb_{\rm 1T}^{\rm gh} \big]^\m{}_\n  =  \d^\m{}_\n \Dz \, ,  \qquad
\big[ \Kbb_{0}^{\rm gh} \big]  = \tfrac{3}{2} \Dz^2 \, , 
\ee
and vertices
\be\label{ghvertex}
\big[ \Mbb_{\rm 1T}^{\rm gh} \big]^\m_\n  = - R^\m_\n \, , \quad
\big[ \Mbb_{0}^{\rm gh} \big]  =  2 D_\m R^{\m\n} D_\n \, , \quad
\left[ \Mbb_{\times}^{\rm gh} \right]^\m = - 2 R^{\m\n} D_\n \, , \quad 
\left[ \Mbb_{\times}^{\rm gh} \right]_\n^\dagger = 2 D^\m R_{\m\n} \, .
\ee
The IR-cutoff following from \eqref{TypeIb} is
\be
\cR_k^{\rm gh,1T} = \d^\m_\n R_k \, , \qquad
\cR_k^{\rm gh,0} = \tfrac{3}{2}\left( P_k^2 - \Dz^2 \right) \, .
\ee

The evaluation of the resulting operator traces then proceeds completely analogous to the previous subsection. The complete ghost-trace decomposes into its $1T$-part
\be\label{ghvector}
\begin{split}
\cS^{\rm gh}_{\rm 1T} = & \, - \Bigg\{
  \Tr_{\rm 1T} \left[ \frac{1}{\Pbb_{\rm 1T}^{\rm gh}} \, \p_t \cR_k^{\rm gh,1T} \right]
  - \Tr_1 \left[ \frac{1}{(\Pbb_{\rm 1T}^{\rm gh})^2} \, \p_t \cR_k^{\rm gh,1T} \Mbb_{\rm 1T}^{\rm gh} \cdot \Pi_{\rm T} \right]
 \Bigg\} \\
= & \, - \tfrac{1}{(4\pi)^2} \int d^4x \sqrt{g} \left\{\,
 3 Q_2[f^1] + R \left( \tfrac{1}{4} Q_1[f^1] + \tfrac{3}{4} Q_2[f^2] \right)
 \,\right\} \, , \\
 \end{split}
 \ee
and the scalar contribution
\be\label{ghscalar}
\begin{split}
\cS^{\rm gh}_0 = & \, - \Bigg\{
  \Tr_0 \left[ \frac{1}{\Pbb_0^{\rm gh}} \, \p_t \cR_k^{\rm gh,0} \right]
  - \Tr_0 \left[ \frac{1}{(\Pbb_0^{\rm gh})^2} \, \p_t \cR_k^{\rm gh,0} \; \Mbb_0^{\rm gh} \right] \Bigg\} \\
= & \, - \tfrac{1}{(4\pi)^2} \int d^4x \sqrt{g} \left\{\,
 2 Q_2[f^1] + R \left( \tfrac{1}{3} Q_1[f^1] + \tfrac{4}{3} Q_3[f^3] \right)
 \,\right\} \, .
\end{split}
\ee
Here, the minus stems from the supertrace for ghost-fields, and the functions $f^{n}$ are given in \eqref{fdef}.
The full contribution of the ghost-sector results from combining \eqref{ghvector} and \eqref{ghscalar}
\be\label{Sghc}
\cS^{\rm gh} =  \, - \tfrac{1}{(4\pi)^2} \int d^4x \sqrt{g} \left\{\,
 5 Q_2[f^1] + R \left( \tfrac{7}{12} Q_1[f^1] + \tfrac{25}{12} Q_3[f^3] \right)
 \,\right\} \, .
\ee
%

%-------------------------------------------------------------
\subsection{Jacobians of the TT-decomposition}
%-------------------------------------------------------------
The final contribution to the universal sector comes from the Jacobians arising from the transverse decomposition of the fluctuation fields. These occur in two types, the Jacobian from the minimal-TT-decomposition and from a vector T-decomposition, $\cS^{\rm aux} = \cS^{\rm aux}_{\rm mTT} + \cS^{\rm aux}_{\rm vec}$. The first one arises from the path-integral over the auxiliary fields given in \eqref{jac:mTT}. Notably, the operator $M_\m^\n$ coincides with the one arising in the ghost sector. Thus
\be\label{SauxmTT}
 \cS^{\rm aux}_{\rm mTT} = \half \cS^{\rm gh} \, ,
\ee
where $\cS^{\rm gh}$ is given by \eqref{Sgh}, and the prefactor takes the combination of bosonic and fermionic vector field into account.

Secondly, we have split two bosonic vectors and two complex fermionic vectors into their transverse and longitudinal components, which in total result in the Jacobian $J_{\rm vec} = [\det{}_{(0)}(\Dz)]^{-1}$. Following (\ref{jac:vec}), this determinant can be exponentiated by introducing a complex scalar field $s, \bar{s}$
\be
\begin{split}
J_{\rm vec} = \int \cD[s] \, {\rm exp}\left[ -\bar{s} \Dz s \right] \, .
\end{split}
\ee
The corresponding contribution to the RG flow is given by a standard scalar trace and easily evaluated to
\be\label{Sauxvec}
\begin{split}
\cS^{\rm aux}_{\rm vec} = \Tr_0 \left[ \frac{1}{P_k}\,\p_t R_k \right] 
= & \tfrac{1}{(4\pi)^2} \int d^4x \sqrt{g}\; \Big\{
Q_2[f^1] + \tfrac{1}{6} R Q_1[f^1] \Big\} \, .
\end{split}
\ee

The trace $\cS^{\rm uni}$ is now found by adding \eqref{Sgftotal}, \eqref{Sghc}, \eqref{SauxmTT}, and \eqref{Sauxvec} and is given in \eqref{tr:uni}.

%-------------------------------------------------------------
\end{appendix}
%-------------------------------------------------------------
%----- Bibliography ----------------------

%-----------------------------------------
%------------------------------------------------------------------------------
\end{document}